\documentclass[%
aps,
prd, 
twocolumn,
superscriptaddress,
nofootinbib,
amsmath,amssymb,amsfonts,
]{revtex4-2}

\usepackage{graphicx, tikz}
\usepackage{bm}
\usepackage[breaklinks=true,colorlinks,citecolor=blue,linkcolor=blue,urlcolor=blue]{hyperref}

\usepackage{mathrsfs}
\usepackage{mathtools}
\usepackage{bbm}

\DeclareMathAlphabet\mathbfscr{OMS}{mdugm}{b}{n}

\DeclareMathSymbol{\shortminus}{\mathbin}{AMSa}{"39}

\usepackage{todonotes}

\renewcommand{\vec}[1]{\boldsymbol{#1}}
\newcommand{\dd}{\mathrm{d}} 
\newcommand{\ee}{\mathrm{e}}
\newcommand{\ii}{\mathrm{i}}

\DeclarePairedDelimiter{\floor}{\lfloor}{\rfloor}
\DeclarePairedDelimiter{\norm}{\lVert}{\rVert}
\newcommand{\avg}[1]{\left\langle {#1} \right\rangle}
\newcommand*{\argdot}{\makebox[1ex]{\textbf{$\cdot$}}}%

\begin{document}

\preprint{APS/123-QED}

\title{Hamiltonian normal forms for the post-Newtonian binary problem}

\author{C. Aykroyd}\email{christopher.aykroyd@obspm.fr}
\affiliation{SYRTE, Observatoire de Paris, PSL Research University, CNRS, Sorbonne Universit\'es, UPMC Univ. Paris 06, LNE, 61 avenue de l'Observatoire, 75014 Paris, France}

\author{A. Bourgoin}
\affiliation{SYRTE, Observatoire de Paris, PSL Research University, CNRS, Sorbonne Universit\'es, UPMC Univ. Paris 06, LNE, 61 avenue de l'Observatoire, 75014 Paris, France}

\author{C. Le Poncin-Lafitte}
\affiliation{SYRTE, Observatoire de Paris, PSL Research University, CNRS, Sorbonne Universit\'es, UPMC Univ. Paris 06, LNE, 61 avenue de l'Observatoire, 75014 Paris, France}

\date{\today}

\begin{abstract}
    We revisit the dynamics of the post-Newtonian (PN) two-body problem for two inspiraling compact bodies. Starting from a matter-only point-mass Hamiltonian, we present an adapted framework based on the Lie series approach, enabling the derivation of complete perturbative solutions within the conservative sector. Our framework supports both circular and eccentric orbits, and is applicable to any perturbation respecting rotational invariance and time-independence. In the context of the Arnowitt–Deser–Misner (ADM) canonical formalism this includes up to at least 3PN order and local terms beyond. We provide an example application at 2PN, recovering classical periapsis advance and orbital period corrections, alongside the full orbital evolution in time coordinates. We discuss eventual extensions to spinning and time-dependent systems.
\end{abstract}

\maketitle


\section{Introduction}

The Einstein field equations form the cornerstone of the general theory of relativity (GR), offering an elegantly profound description of interacting matter, energy, and geometry. Yet these equations are notoriously complex due to their nonlinear, tensorial and interdependent nature. As a consequence, the discovery of exact solutions is chiefly limited to highly idealized symmetrical scenarios. Real astrophysical phenomena, however, are rarely encapsulated by such pristine conditions. 
Gravitationally radiating systems are a prime example of inherently asymmetric systems, particularly evident in the dynamics of compact binaries. Their study typically segments the process into distinct regimes, each requiring its own set of numerical or analytical approximation tools. The PN framework, popularized from the earliest days of GR \cite{Einstein1916, Lorentz1937, Levi-Civita1937, Einstein1938}, excels in addressing the inspiralling two-body problem. Initially, PN approximations assume wide, bound orbits with weakly or moderately relativistic speeds. Yet in practice, these solutions have been `unreasonably effective' \cite{Will2011} in describing a range of compact systems, achieving remarkable accuracy up to the innermost stable circular orbit. This success is partly attributed to GR's `effacing' property, which permits the omission of self-gravitation and internal structure at lower orders.

The evolution of the PN framework has closely followed technological advancements, notably with the discovery of the Hulse-Taylor pulsar \cite{Hulse1975}, which underscored the necessity for a formalism compatible with high-order calculations. The era of gravitational wave (GW) astronomy, inaugurated by the detection of GW150914 \cite{Abbott2016}, has reinforced this need, aligning analytical methods with the capabilities of second-generation detectors such as Advanced LIGO \cite{LIGO}, Virgo \cite{VIRGO}, KAGRA \cite{KAGRA} (LVK), and the International Pulsar Timing Array \cite{IPTA}. As we approach the advent of third-generation detectors, including LISA \cite{LISA}, the Einstein Telescope \cite{ET}, and the Cosmic Explorer \cite{CosmicExplorer}, further refinement of the PN framework remains imperative. 

PN schemes typically rely on decoupling the field and matter degrees-of-freedom.
This can be performed by `integrating out' the metric either at the level of the field equations or directly at the action---via a combination of perturbative techniques such as multipolar expansions \cite{Ross2012}, asymptotic matching \cite{Blanchet2024, Poisson2014}, and Effective Field Theory \cite{Goldberger2006, Porto2016, Levi2020}. Eventually, the aim is to establish a matter-only Fokker Lagrangian \cite{Henry2023}, or to directly compute the particles' equations of motion \cite{Blanchet2024}.
Canonical formulations like ADM instead search for a matter-only Hamiltonian \cite{Jaranowski1998, Schafer2024}. Accordingly, different harmonic and ADM-TT gauge formulations have been found to be in agreement \cite{Andrade2001, Blanchet2024}. 
But while there has been a notable focus in this field-matter decoupling step of PN schemes, there remains a substantial gap in furnishing comprehensive analytical solutions of the resulting functionals. The quasi-Keplerian parametrisation \cite{Damour1985, Memmesheimer2004, Cho2022} has been sucessfully determined up to 4PN to describe the orbits of non-spinning binaries. However,  a systematic extension of this approach to broader perturbative situations including spinning or dissipative systems remains a challenge. 
In many other instances, orbital evolutions are only determined approximately, restricted to secular contributions or assumptions of (quasi-)circular orbits. A common approach employs the Lagrange Planetary Equations \cite{Lincoln1990, Tucker2021}, which, fundamentally, entails solving by hand the equations of motion for the Keplerian elements, and can hardly be achieved exactly. 
In the context of canonical formalisms, Hamilton-Jacobi methods \cite{Damour2000a, LeTiec2015, Schafer2024}---typically represented in action-angle coordinates---provide a straighter path to determining long-term dynamical invariants, but often lack explicit time-coordinate solutions.

Departing from the canonical formalism for point-mass binaries, the Lie series approach can instead allow us to achieve an analytic portrayal of the dynamics that captures both long- and the short-term phenomena. Notably, the Lie framework provides a very structural path for extending `secular averaging' procedures to high orders and to a broad range of perturbative scenarios. The approach has been applied to solve the binary problem at 1PN order, in the spinless case \cite{Richardson1988} and in the spinning case \cite{Biscani2012, Biscani2014}. The next-to-leading order spin-orbit \emph{contributions} to point-mass 2PN circular orbits have been computed in \cite{Tessmer2013}.
To our knowledge, a general two-body Lie framework that is directly applicable \emph{for solving the motion} of post-Newtonian binaries (at higher orders) is not yet established.
More broadly, Lie-series techniques have also been employed for determining symplectic coordinate transformations to the centre-of-mass frame in the ADM formalism \cite{Rothe2010}; or between harmonic, ADM, effective one-body and isotropic coordinates \cite{Bluemlein2020a, Bluemlein2020b}.

In this work, we aim to describe a canonical framework for solving the two-body dynamics that can be readily applied to a wide range of perturbed systems.
We accordingly furnish parametric solutions for time-independent and rotationally-invariant perturbations. These assumptions are valid in the ADM (spinless) conservative sector, up to at least 3PN order and to local terms beyond. Our provided solutions shall include the complete analytical orbital behaviour at some PN truncation order, including secular and oscillatory terms, and valid for fully eccentric systems.
We shall prescribe a treatment which does not require explicit usage of action-angle coordinates, but which is instead established in terms of Keplerian functions or other parametrisations that can be easily adapted to the problem at hand. This freedom facilitates eventual applications of the framework to spinning systems and perturbations of non-gravitational nature.
In Sec.~\ref{sec:ADM}, we shall apply our framework to the 2PN ADM-TT gauge Hamiltonian, not only re-deriving standard periapsis advance and orbital period corrections, but also thoroughly exploring the full long- and short-term orbital evolution. We then verify our results in a numerical example.

\paragraph*{Notation and conventions.} \phantom{\null}\par

In this paper, we shall denote by $G$ the gravitational constant and by $c$ the speed of light.

Complex conjugation is represented by the dagger operator $(\cdot)^\dag$. Bold symbols are reserved for 3-vectors, decomposed into norm $u = \norm{\vec u}$ (light-typeface) and direction $\hat{\vec u} = \vec u / u$ (overhat).

We adopt the following convention for the Poisson bracket between two functions $f$ and $g$ on phase-space with coordinates $(\boldsymbol{r}, \boldsymbol{p})$:
\begin{align}
     \{ f, g \} \coloneqq \frac{\partial f}{\partial \vec r} \cdot \frac{\partial g}{\partial \vec p} - \frac{\partial f}{\partial \vec p} \cdot \frac{\partial g}{\partial \vec r} \text,
\end{align}
with the standard Euclidean dot product.

Thus, two functions $f$ and $g$ are said to be in \emph{involution} when
\begin{equation}
    \{ f, g \} = 0 \text.
\end{equation}

We adopt the Liouville sense of integrability. A Hamiltonian system with phase-space dimension $2N$ is said to be \emph{integrable} if it admits at least $N$ independent integrals of motion $\vec I = (I_1, \ldots I_N)$ which are in mutual involution---that is, for all pairs $(I_n, I_m)$ elements of $\vec I$:
\begin{equation}
    \{ I_n, I_m \} = 0 \text.
\end{equation}
Integrable systems are desirable because they can in principle be solved by quadratures (for instance in action-angle variables) \cite{Arnold2013}.
The system is further said to be \emph{maximally superintegrable} when it admits $2N-1$ independent integrals of motion in involution. In this case, the system's trajectory is completely determined by these constants and can, in principle, even be solved algebraically (provided that the necessary inversions can be performed).

The \emph{flow} $\Phi_t^{\mathcal H}$ generated by a Hamiltonian $\mathcal H$ is a one-parameter group of canonical transformations which maps out the evolution of any point in phase-space from time $t_0$ to time $t_0 + t$:
\begin{equation}
    \Phi_t^{\mathcal H} (\vec r(t_0), \vec p(t_0)) = (\vec r(t_0 + t), \vec p(t_0 + t)) \text,
\end{equation}
where $(\vec r, \vec p)$ evolve according to the Hamiltonian equations generated by $\mathcal H$. 

\section{Two-body perturbative framework}
\label{sec:framework}
We dedicate this section to recalling the general Lie perturbative method and formulating it in the context of the post-Newtonian two-body problem.
Consider a local, autonomous Hamiltonian describing the matter dynamics for two point-particles of the form:
\begin{equation}
    \mathcal H(\vec r, \vec p)  = \sum_{\ell=0}^{K-1} c^{-2\ell} \mathcal H_{\ell}(\vec r, \vec p) + \mathcal O(c^{-2K}) \text, \label{eq:Hamiltonian}
\end{equation}
expressed in a center-of-mass coordinate system $\mathscr C$. 
Here, $\mathcal H_0$ is the zeroth-order Newtonian term, assumed integrable, and $c^{-2\ell} \mathcal H_\ell(\vec r, \vec p)$ are the PN perturbations at order $\ell \geq 1$.
We treat $c$ as a bookkeeping parameter, with the order of terms denoted using $\mathcal O(c^{-2\ell})$.
More formally, the expansion can be characterized by a small dimensionless parameter, 
\begin{equation}
    \varepsilon \sim \max{ \left( \frac{v}{c}, \sqrt{\frac{G m}{r c^2}} \right) } \text.
\end{equation}
where $v$ is a characteristic velocity and $m = m_1 + m_2$ represents the total mass parameter of the binary.
Each Hamiltonian term is therefore assumed to be bounded as $c^{-2\ell} \lvert \mathcal H_\ell(\vec r, \vec p) \rvert \leq \kappa \varepsilon^{2\ell}$ for some $\kappa > 0$ and small enough $\varepsilon$.

Since the non-integrable perturbations enter the Hamiltonian at higher than first order, the system is called quasi-integrable. Fortunately, despite the non-integrability of individual terms, such systems often allow perturbative solutions. In the Lie approach, these solutions are established via an appropriate change of coordinates.

The ideal goal of the method is to construct a canonical transformation to a new coordinate system $\mathscr C^*$ in which the transformed Hamiltonian $\mathcal H^*$ becomes integrable:
\begin{align}
    \mathcal H^*(\vec r^*, \vec p^*) ={}& \sum_{\ell=0}^{K-1} c^{-2\ell} \mathcal H^*_\ell(\vec r^*, \vec p^*) + \mathcal O(c^{-2K}) \text.\label{eq:Hamiltonian_normal_form}     
\end{align}
where the starred variables $\vec r^*$ and $\vec p^*$ are the new canonical variables in $\mathscr C^*$. The Hamiltonian terms $\mathcal H^*_\ell$ are now in mutual involution\footnote{This can be seen by remarking that they are functions of the action variables only, and any two functions of action variables are always in involution.}, which ensures the integrability of $\mathcal H^*$. Effectively, any non-integrable terms have been pushed to orders higher than the truncation order $K - 1$. The above expression is known as a \emph{Birkhoff normal form}.

Unfortunately, in general, the Lie procedure cannot be carried out to infinite order, but there exists an optimal order $Q \leq K-1$ at which the truncation residuals are minimised\footnote{Rigorous estimates of the residuals and the convergence radius of the Hamiltonian normal form can be found, for example, in~\cite{Giorgilli1985}.}. More generally, the normal form can be written as:
\begin{align}
    \mathcal H^*(\vec r^*, \vec p^*) ={}& \sum_{\ell=0}^Q c^{-2\ell} \mathcal H^*_\ell(\vec r^*, \vec p^*) \nonumber\\
    &{} + \sum_{\ell=0}^{K-1} c^{-2\ell} \mathcal R_\ell(\vec r^*, \vec p^*) + \mathcal O(c^{-2K}) \text,\label{eq:Hamiltonian_normal_form_with_residuals}     
\end{align}
where the terms $\mathcal R_\ell$ are non-integrable remainders which could not be eliminated in the procedure: these include non-integrable terms from orders higher than the optimal order ($\ell > Q$), as well as eventual resonances from the lower orders ($\ell \leq Q$); when present, such resonances must be carefully handled (see e.g.\ \cite{Morbidelli2002} for details).

For the class of systems we shall be considering in this paper, each Hamiltonian term shall be a rotationally-invariant scalar, depending only on $r = \norm{\vec r}$, $p_r \coloneqq \vec p \cdot \vec r / r$ and $p = \norm{\vec p}$. We shall see \textit{a posteriori} that resonances which could complicate the normal form are absent in our systems---thus, $\mathcal R_\ell = 0, \forall \ell \leq Q$. 
Furthermore, we shall henceforth restrict $\mathcal H$ to be of PN order sufficiently low that the optimal order exceeds the truncation order ($Q \geq K - 1$). In this case, the normal form\ \eqref{eq:Hamiltonian_normal_form_with_residuals} is reduced to \eqref{eq:Hamiltonian_normal_form}.

The general strategy to obtain \eqref{eq:Hamiltonian_normal_form} or \eqref{eq:Hamiltonian_normal_form_with_residuals} is to consider a family of `near-identity' coordinate transformations which are canonical by construction. Namely, we consider the (infinite-dimensional) group of Hamiltonian symplectomorphisms, coordinate transformations $\mathcal T_g$ generated by the action of smooth functions $g$. Equipped with the Poisson bracket these functions form a Lie algebra with elements $\mathfrak L_g = \left\{ \cdot , g \right\}$; each vector element may act infinitesimally on phase-space functions $f$:
\begin{equation}
     \delta f = \mathfrak L_g(f) = \left\{ f, g \right\} \text.
\end{equation}
The near-identity condition requires the generator to be of PN order $\mathcal O(c^{-2})$, with potentially higher-order terms:
\begin{equation}
    g(\vec r, \vec p) = \sum_{\ell=1}^{K-1} c^{-2\ell} g_\ell(\vec r, \vec p) + \mathcal O(c^{-2K}) \text,
\end{equation}
where each $g_\ell$ is explicitly independent of $c$.
To recover the global coordinate transformation one then exponentiates $\mathfrak L_g$, giving rise to the \emph{Lie series}:
\begin{equation}
\begin{aligned}
    \mathcal T_g(f) ={}& \ee^{\mathfrak L_g} (f) \\ \coloneqq{}& \sum_{\ell=0}^\infty \frac{1}{\ell!} \mathfrak L_g^{\ell}(f) \\ ={}& 
    f + \{ f , g \} + \frac{1}{2} \big\{ \{ f , g \} , g \big\} + \ldots \text,    
\end{aligned}    
\end{equation}
where $\mathfrak L_g^{\ell}$ denotes the $\ell$-th repeated composition of $\mathfrak L_g$. By construction, these transformations preserve Poisson brackets, since they represent the flow generated by $g$ parametrised by some variable $\lambda$:
\begin{equation}
\mathcal T_g = \Phi^g_{\lambda = 1} \text. 
\end{equation}
Note that the converse is not true, not all symplectic transformations can be expressed as the application of a Hamiltonian flow.
Under this formalism, the phase-space coordinates transform as:
\begin{align}
    \vec r = \mathcal T_{g}(\vec r^*)\text, && \vec p = \mathcal T_{g}(\vec p^*)\text,
    \label{eq:phase_space_transformation}
\end{align}
with the inverse given by negating the sign of $g$.

\subsection{Homologic equations}
Time-independent canonical transformations preserve the value of the Hamiltonian, in which case it holds that $\mathcal H^*(\vec r^*, \vec p^*) = \mathcal H(\vec r, \vec p) = \mathcal T_g (\mathcal H)(\vec r^*, \vec p^*)$. Expanding the Lie operator on $\mathcal H$ and equating similar-order terms yields the \textit{homologic} functional relations:
\begin{subequations}
\label{eq:homologic_equation_system}
\begin{align}
    \Big.\mathcal H_0 & = \mathcal H^*_0 \\
    \big\{ g_1, \mathcal H_0 \big\} & = \mathcal H_1 - \mathcal H^*_1 \text, \\
    \big\{ g_2, \mathcal H_0 \big\} & = \mathcal H_2 + \frac{1}{2} \big\{ \mathcal H_1 + \mathcal H^*_1, g_1\big\} - \mathcal H^*_2 \text, \\
    & \ \vdots \nonumber
\end{align}
\end{subequations}
where the unknowns are the generator sequence $(g_k)$ and the Hamiltonian sequence $(\mathcal H^*_k)$.
The homologic equations can be solved sequentially from the top, plugging the solutions from the previous iterations; generically, they have the form: 
\begin{equation}
    \big\{ g_\ell, \mathcal H_0 \big\} = \mathcal P_\ell - \mathcal H^*_\ell \text,
    \label{eq:homologic_equation}
\end{equation}
where the perturbation $\mathcal P_\ell$ is some function of the solutions up to order $\ell-1$.
Taking the average along the Newtonian flow $\Phi^{\mathcal H_0}_t$ directly yields a solution for $\mathcal H^*_\ell$:
\begin{align}
    \mathcal H^*_\ell &= \avg{\mathcal P_\ell} \text, \label{eq:Hamiltonian_solution}
\end{align}
where
\begin{equation}
    \avg{f} = \lim_{T \to \infty} \frac{1}{T}\int_{-T/2}^{T/2} f \circ \Phi^{\mathcal H_0}_t \dd t \text.
\end{equation}

Next, we determine the generator by leveraging the integrability of the Keplerian system.
Namely, $\mathcal H_0$ is a maximal superintegrable system of $6$-dimensional phase-space with $5$ independent integrals of motion. These integrals can be specified in many forms; 
we shall supply some of them in the following section.
In this manner, the system can be expressed in terms of five freely-chosen independent constants $\vec{I} = (I_1, \ldots I_5)$ in involution,
and a remaining Keplerian periodic degree-of-freedom, which we shall call $\theta$. We stress that although these six parameters must be one-to-one with $(\vec r, \vec p)$, they are not required to form a canonical set, allowing more flexibility in the specification of the results\footnote{For instance, formulating the problem uniquely in terms of Delaunay variables might force one to expand equations in series of eccentricity. Furthermore, the freedom of choice facilitates the extension of the method to non-canonical degrees-of-freedom such as spins.}.
We may subsequently develop the left-side of the homologic equation via chain rule:
\begin{equation}
    \{ g_\ell, \mathcal H_0 \} =  \frac{\partial g_\ell}{\partial \theta} \{ \theta, \mathcal H_0 \} + \frac{\partial g_\ell}{\partial \vec{I}} \cdot \{ \vec{I}, \mathcal H_0 \} \text,
\end{equation}
where the five brackets $\{ \vec{I}, \mathcal H_0\}$ identically vanish by definition.
A formal solution is thus obtained from Eq.\ \eqref{eq:homologic_equation} by integrating over $\theta$:
\begin{align}
    g_\ell &= \int \big(\mathcal P_\ell - \mathcal H^*_\ell \big) \left( \frac{\dd \theta}{\dd t}\right)^{-1}_{\mathrm{kep}} \,\dd \theta \text,
    \label{eq:generator_solution}
\end{align}
where we have defined the Keplerian time-derivative
\begin{equation}
    \left(\frac{\dd }{\dd t}\right)_{\mathrlap{\mathrm{kep}}} \quad \equiv \{ \argdot, \mathcal H_0 \}   \text.
\end{equation}
Some care must be taken so that $\{ \theta, \mathcal H_0 \} \neq 0$.
The above primitive is taken to be average-free, which avoids the emergence of Poisson terms.

The construction of an explicit parametric expansion to these integral solutions requires a convenient description of the system coordinates. For this we propose the Keplerian parametrisation of the following section.

\subsection{The Keplerian parametrisation}
\label{sec:Keplerian_parametrisation}

We shall now recall the Keplerian osculating parameters, which are typically employed in celestial mechanics (see e.g.\ \cite{Murray2000}) and are convenient for the Lie formalism.
We stress that this differs from the \emph{post-Keplerian} formalism \cite{Damour1985, Memmesheimer2004}, where Newtonian relationships are adjusted with PN corrections. Instead, one adopts phase-space relations which are exact, independently of the system perturbation. Namely, every given point $(\vec r, \vec p)$ in phase-space $\mathscr C$ (or likewise in $\mathscr C^*$) is uniquely characterised by six orbital elements; for instance: the semi-major axis $a$, the eccentricity $e$, the true anomaly $v$, the argument of the periapsis $\omega$, the inclination $\iota$, and the longitude of the node $\Omega$. These do not form a canonical set of variables, yet their Poisson brackets do not require PN corrections, facilitating the integration of the Newtonian-level derivatives in the homologic equation. We shall also introduce the eccentric and mean anomalies $E$ and $M$, the longitude of the periapsis $\varpi = \omega + \Omega$, and the total orbital action $L = \sqrt{a}$, all of which can be related to the aforementioned six elements. To specify all these parameters, we henceforth assume that the coordinates are appropriately rescaled, so that the Keplerian potential adopts the form:
\begin{equation}
    \mathcal H_0(\vec p, \vec r) \equiv \frac{p^2}{2} - \frac{1}{r} = - \frac{1}{2 L^2} \text.
    \label{eq:Hamiltonian_Kepler}
\end{equation}
At each point in $\mathscr C$, we introduce an ellipse tangent to the motion, lying within a coordinate plane perpendicular to the orbital angular momentum $\vec{J}(\vec r, \vec p) = \vec r \times \vec p$. Accordingly, we decompose the motion onto a right-handed orbital basis $(\hat{\vec A},\hat{\vec B},\hat{\vec J})$:
\begin{subequations}
\begin{align}
    \vec r &= r \, ( \cos v \, \hat{\vec A} + \sin v \, \hat{\vec B} ) \text, \\
    \vec p &= \frac{1}{J} \big( - \sin v \, \hat{\vec A} + (e + \cos v) \, \hat{\vec B} \big) \text,
\end{align}    
\end{subequations}
with $r = J^2 / (1 + e \cos v)$ and $e = \sqrt{1 - J^2/L^2}$. The vectors $\vec{A} = e \hat{\vec A}$ and $\vec{B} = e \hat{\vec B}$ are the periapsis and binormal vectors. Together with $\vec J = J \hat{\vec J}$ they form nine (dependent) first integrals of the Newtonian motion, of which $\vec{A}$ and $\vec{B}$ are only conserved for perfect inverse-square central forces. The last two can be readily expressed in functional form:
\begin{align}
    \vec{A} = \vec p \times \vec{J} - \frac{\vec r}{r}
    \text, &&
    \vec{B} = J \vec p - \frac{\vec {J} \times \vec r}{J r} \text.
\end{align}
The orbital basis can be further related to a non-rotating reference orthonormal basis $(\hat{\vec X}, \hat{\vec Y}, \hat{\vec Z})$ via a rotation operator $R$:
\begin{align}
    \hat{\vec A} = R (\hat{\vec X}) \text, && 
    \hat{\vec B} = R (\hat{\vec Y}) \text, && 
    \hat{\vec J} = R (\hat{\vec Z}) \text.
\end{align}
It is typical (but not required) to specify $R$ via Euler angles $(\omega, \iota, \Omega)$ following the $Z${-}$X${-}$Z$ convention\footnote{In the $Z${-}$X${-}$Z$ extrinsic convention, an arbitrary rotation is decomposed as a sequence of three right-handed rotations about the specified inertial axes $\hat{\vec Z}$, $\hat{\vec X}$, and $\hat{\vec Z}$, in this order. The respective rotation angles are $\omega$, $\iota$, and $\Omega$. The complete rotation operator is thus constructed from the individual axis rotations as $R = R^{\hat Z}_\Omega \circ R^{\hat X}_\iota \circ R^{\hat Z}_\omega$.}.

Finally, the three anomalies $v$, $E$ and $M$ are defined to satisfy the following relationships:
\begin{equation}
    M = E - e \sin E \label{eq:mean_anomaly}
\end{equation}
and
\begin{subequations}\label{eq:true_and_eccentric_anomalies}
\begin{align}
    \cos v &= \frac{1}{e} \left( \frac{J^2}{r} - 1 \right) \text, & \sin v &= \frac{J p_r}{e} \text, \\
    \cos E &= \frac{1}{e} \left(1 - \frac{r}{L^2} \right) \text, & \sin E &= \frac{r p_r}{e L} \text.
\end{align} 
\end{subequations}
These angles are chosen such that their time evolution is continuous, with the difference between any two of them not exceeding $\pi$. This is achieved by selecting the appropriate branches of the inverse trigonometric functions when solving Eqs.\ (\ref{eq:mean_anomaly}, \ref{eq:true_and_eccentric_anomalies}), adding multiples of $2\pi$ as necessary to maintain time continuity.


We emphasize that the Keplerian parameters are generally to be seen as functions of phase-space and applicable to both the $\mathscr C$ and $\mathscr C^*$ coordinate systems. Accordingly,  for each parameter $x$, we introduce their time-evolutions under Hamiltonian $\mathcal H$ and $\mathcal H^*$:

\begin{subequations}
    \begin{align}
        \tilde x(t) &= x(\vec r (t), \vec p(t)) \text, \\
        \bar x(t) &= x(\vec r^* (t), \vec p^*(t)) \text.
    \end{align}
\label{eq:convention_time_dependence}
\end{subequations}
We shall refer to $\bar x$ as the secular evolution of $x$ and to $\tilde x$ as the complete time evolution of $x$. We can immediately deduce the following relationship:

\begin{equation}
    \tilde x = \overline{\mathcal T_{g} (x)} \text.
    \label{eq:Keplerian_parameter_transformation}
\end{equation}
The evolution equation of each parameter can be extracted from the generalized Hamilton's equations,

\begin{subequations}
\begin{align}
    \dot{\tilde x} &= \widetilde { \{ x, \mathcal H \} } \text, \label{eq:hamilton_equations_tilde} \\
    \dot{\bar x} &= \overline { \{ x, \mathcal H^* \} } \text, \label{eq:hamilton_equations_bar}
\end{align}
\end{subequations}
valid for non-canonical variables. In these equations, the application of \emph{tilde} (resp.\ \emph{bar}) denote that expressions are evaluated in $\mathscr C$ (resp.\ $\mathscr C^*$) coordinates [cf.\ Eq.\ \eqref{eq:convention_time_dependence}], after computation of the brackets. 
The advantage of the Lie approach lies in that solving Eq.\ \eqref{eq:hamilton_equations_bar} and then applying \eqref{eq:Keplerian_parameter_transformation} is in general much easier than directly solving \eqref{eq:hamilton_equations_tilde}.
The Poisson bracket structure of each parameter $x$ can be deduced from the provided definitions, and is already known from celestial mechanics\footnote{For instance, the Delaunay set $(M, \omega, \Omega, L, J, H = J \cos \iota)$ is conveniently canonical with $\{M, L\} = \{\omega, J\} = \{ \Omega, H \} = 1$ and all other bracket combinations equal to zero.}. 
We remark that these brackets remain unchanged under the symplectic transformation $\mathcal T_g$.
 
\subsection{Solutions to the homologic equation}
The integral solutions to the homologic equation can be explicitly solved for a large class of rotationally-invariant problems. Namely, we consider the following polynomial expansion in $1/r$: 
\begin{equation}
    \mathcal P_\ell  = A_{\ell, 0} + A_{\ell, 1} r^{-1} + \ldots + A_{\ell, s} r^{-s} \text, \label{eq:polynomial_r_expansion}
\end{equation} 
expressed such that the coefficients $A_{\ell, 0},\ldots,A_{\ell, s}$ depend solely on the Keplerian integrals of motion $\vec{I}$ (via use of the relationships from Sec.~\ref{sec:Keplerian_parametrisation}). In this case, we develop Eqs.~(\ref{eq:Hamiltonian_solution}, \ref{eq:generator_solution}) with:
\begin{align}
    g_\ell &= \sum_{k=1}^s A_{\ell, k} \left(\xi_k - \frac{M}{n_0} h_k \right) \text, &
    \mathcal H^*_\ell &= \sum_{k=0}^s A_{\ell, k} h_k \text,\label{eq:homologic_solution_iterative}
\end{align}
with $n_0 = (- 2 \mathcal H_0)^{3/2}$.
The basis elements turn out to be:
\begin{align}
    \xi_k = \int \frac{1}{r^k} \left( \frac{\dd \theta}{\dd t}\right)^{-1}_{\mathrm{kep}} \,\dd \theta \text, && h_k = \avg{ \frac{1}{r^k} }\text,
\end{align}
where the primitive is chosen ensuring that $\avg{\xi_k} = (\avg{M}/n_0) h_k = 0$. Explicit integrated expressions for $\xi_k$ and $h_k$ are provided in Appendix~\ref{appendix:flow_integrals}.

With this the Hamiltonian $\mathcal H^* = \sum_\ell c^{-2\ell} \mathcal H^*_\ell$ and the generator $g = \sum_\ell c^{-2\ell} g_\ell$ are now fully determined. The explicit coordinate-time evolutions $\vec r^*(t)$ and $\vec p^*(t)$ can hence be obtained from the \emph{integrable} equations of motion; for instance, via the Kepler elements $\bar x(t)$ outlined in the previous section. Initial conditions in $\mathscr C^*$-space are determined through application of the inverse Lie operator $\mathcal T_{-g}$.
Eventually, the phase-space evolution in the original coordinate system $\mathscr C$ can be recovered via application of $\mathcal T_g$. 

As an immediate example, we shall apply the framework to the conservative sector of the ADM matter Hamiltonian at 2PN order.

\section{Application to the canonical ADM Hamiltonian at 2PN}
\label{sec:ADM}

Asymptotically flat coordinates in GR admit the Poincar\'e group as a global symmetry \cite{Andrade2001,LeTiec2015}. The generators of this group are realised as 10 surface integrals, conserved on-shell as dictated by Noether's theorem: the total energy $\mathscr E$ and total linear momentum $\mathbfscr{P}$, generators of space-time translations; the angular momentum $\mathbfscr{J}$, generator of rotations; and the generator of boosts $\mathbfscr{K} = \mathbfscr{G} - t \mathbfscr{P}$ associated to the centre-of-mass constant $\mathbfscr{G}$. In canonical formalisms, each of these quantities is seen as a function of phase-space. The ADM-TT gauge Hamiltonian is not \emph{manifestly} Poincar\'e invariant since most quantities are realised in nonlinear fashion. Instead, this invariance must be explicitly shown as done in \cite{Damour1981, Damour1985, Damour2000, Damour2008, Schafer2024}.
However, for non-spinning particles, the ADM-TT gauge does manifestly respect the \emph{Euclidean} group, and thus the momenta are simply $\mathbfscr{P} = \sum_a \vec p_a$ and $\mathbfscr{J} = \sum_a \vec r_a \times \vec p_a$, valid at all post-Newtonian orders (where $(\vec r_a, \vec p_a)$ are the coordinates of particle $a$). The energy, as usual, refers to the conserved Hamiltonian, namely $\mathscr E = \mathcal H - mc^2$.
When including the radiative sector to account for losses via the gravitational radiative reaction, the Noether quantities are no longer conserved; at orders $2.5$PN and above, one must consider flux-balance-type equations (see e.g.\ \cite{Andrade2001, Compere2020, Henry2023}) or time-dependent Hamiltonians \cite{Koenigsdoerffer2003, Schafer2024}.
In our case we are concerned in studying the conservative sector of the two-body problem.

From the freedom of performing a Poincar\'e transformation, we adopt coordinates of the centre-of-mass (CM) where $\mathbfscr{P} = \mathbfscr{G} = \mathbfscr{K} = 0$.
Rothe and Sch\"afer \cite{Rothe2010} explicitly determine a symplectic transformation to a canonical CM frame which is valid for any $\mathbfscr{P}$ and $\mathbfscr{G}$. In practice, under CM coordinate conditions ($\mathbfscr{P} = \mathbfscr{G} = \mathbfscr{K} = 0$), the frame transformation reduces to the Newtonian-like relations $\vec p = \vec p_1 = - \vec p_2$, $\vec r = \vec r_2 - \vec r_1$ (see also \cite{Damour1988, Jaranowski1998, Damour2000a}).
As our starting point, we shall adopt the rescaled CM coordinates:
\begin{align}
    \vec p' &= \frac{\vec p}{\mu} \text,
    & \vec r' &= \frac{\vec r}{G m} \text, 
    & \mathcal H' &= \frac{\mathcal H - m c^2}{\mu} \text,
    & t' &= \frac{t}{G m} \text, 
\end{align} where $\mu = m_1 m_2 / m$ and $\nu = \mu / m$. We also rescale $\mathscr E' = \mathscr E/\mu$ and $\mathbfscr J' = \mathbfscr J / (G m \mu)$. In these coordinates, the small parameter scales as $\varepsilon' \sim (1/c)\max{(v, \sqrt{1/r})}$.
For notational simplicity, we shall henceforth omit the primes. In rescaled coordinates, the point-mass ADM-TT Hamiltonian reads \cite{Damour2000a, Schafer2024}:
\begin{subequations}
\label{eq:Hamiltonian_ADM}
\begin{align}
    \mathcal H_1(\vec r, \vec p) ={}& \frac{1}{2 r^2} - \frac{1}{8} (1 - 3\nu) p^4  \nonumber\\&
    - \frac{1}{2 r} \Big( (3 + \nu) p^2 + \nu (\vec n \cdot \vec p)^2 \Big)  \text, \\
    \mathcal H_2(\vec r, \vec p) ={}& \frac{1}{16} \big( 1 - 5 \nu + 5 \nu^2 \big) p^6 + \frac{1}{8 r} 
    \Big( \!\left( 5 - 20\nu - 3\nu^2 \right) \! p^4 \nonumber\\&
    - 2 \nu^2 p^2 (\vec n \cdot \vec p)^2  - 3 \nu^2 (\vec n \cdot \vec p)^4 \Big) \nonumber\\&
    + \frac{1}{2 r^2} \Big( (5 + 8\nu) p^2 + 3\nu (\vec n \cdot \vec p)^2 \Big) \nonumber\\&
    - \frac{1}{4 r^3} (1+3\nu) \text,
\end{align} 
\end{subequations}
with $\vec n =\vec r/r$ and leading zeroth-order term $\mathcal H_0$ as specified in Eq.~\eqref{eq:Hamiltonian_Kepler}.

\vspace{1.5em} 

\subsection{Solution and observables}
We employ the methods outlined in Sec.~\ref{sec:framework} to transform the rescaled CM Hamiltonian. Specifically, Eq.~\eqref{eq:homologic_solution_iterative} parametrically furnishes the normal-form and generator as a function of the perturbation at each PN order.
We consider iteratively the two perturbations: 
\begin{align}
    \mathcal P_1 = \mathcal H_1 \text, && \mathcal P_2 = \mathcal H_2 + \frac{1}{2} \{ \mathcal H_1 + \mathcal H_1^*, g_1 \} \text.
\end{align}
Each perturbation is then expanded into the form \eqref{eq:polynomial_r_expansion} by appropriately expressing momenta in terms of the first integrals and the separation, with the aid of the following two expressions:
    \begin{align}
        p^2 = \frac{2}{r} - \frac{1}{L^2} \text, && 
        p_r^2 = p^2 - \frac{J^2}{r^2} \text.\label{eq:polynomial_r_subs}
    \end{align}
The procedure leads to the following integrable 2PN Hamiltonian:
\begin{align}
    {\mathcal H}^* (J,L) =&  - \frac{1}{2 L^2} + \frac{1}{c^2} \bigg\{ \frac{15 - \nu}{8 L^4} - \frac{3}{J L^3} \bigg\} \nonumber\\& 
    + \frac{1}{c^4} \bigg\{  \frac{5 (2 \nu - 7)}{4 J^3 L^3}-\frac{27}{2J^2 L^4} + \frac{3 (35 - 4 \nu)}{4 J L^5} \nonumber\\& 
    - \frac{\nu ^2-15 \nu +145}{16 L^6} \bigg\} \text. \label{eq:Hamiltonian_ADM_normal_form}
\end{align}
We recall that the parameters $L = L(\vec r^*, \vec p^*)$ and $J = J(\vec r^*, \vec p^*)$ are treated as pure Newtonian-order functions of phase-space coordinates. This enables calculations to be treated agnostically in terms of the parameter choice and more readily adapted to a new problem.
Simultaneously, Hamiltonian~\eqref{eq:Hamiltonian_ADM_normal_form} can also be regarded as if it were expressed in Delaunay variables, since the involved transformation is canonical.
Unsurprisingly, we find that our expression precisely matches the Delaunay form that can be derived via Hamilton-Jacobi methods (cf.~\cite{Damour2000a, Schafer2024}). Nevertheless, Lie methods have the advantage of also providing the generator:
\begin{widetext}
\begin{align}
    g =& \frac{1}{c^2} \left\{ \frac{p_r r (J (4 - \nu) - 6 L)}{2 J L^2}-\frac{6}{J} \arctan{\!\left(\frac{L r p_r}{J L+r}\right)} +\frac{\nu  p_r}{2} \right\}
    + \frac{1}{c^4} \Bigg\{ 
    \frac{p_r}{r^2} \bigg( \frac{J^2 (\nu^2 -2 \nu)}{8} -\frac{3 J^3 \nu }{4 (J+L)} \bigg) \nonumber\\&
    + \frac{p_r}{r} \left( \frac{9 J}{2 (J+L)} + \frac{6 -\nu}{4} \right) 
    + p_r \bigg( \frac{5 (2 \nu -7)}{4 J^2}+\frac{3 (\nu +2)}{4 J (J+L)}+\frac{3}{J L} +\frac{-\nu ^2+10 \nu -32}{8 L^2} \bigg)  \nonumber\\& 
    + p_r r \left(\frac{5 (2 \nu -7)}{4 J^3 L}+\frac{9 (3 - \nu)}{4 J L^3}-\frac{\nu+16}{8L^4}\right)
    +\left(\frac{10 \nu - 35}{2 J^3}+\frac{15-6 \nu }{2 J L^2}\right) \arctan\left(\frac{L r p_r}{J L+r}\right)
    \Bigg\} \text. \label{eq:generator_ADM}
\end{align}
\end{widetext}

We have verified the consistency of the obtained generator, checking that our expressions for $\mathcal H^*$ and $g$ identically satisfy the homologic equations---to order $\mathcal O(c^{-6})$---with $\mathcal T_g$ correctly transforming $\mathcal H$ into $\mathcal H^*$. In the next steps, we shall derive the integrable dynamics from Hamiltonian \eqref{eq:Hamiltonian_ADM_normal_form}, and eventually incorporate the short-term oscillatory behaviour to obtain the complete dynamics via application of the Lie transform generated by \eqref{eq:generator_ADM}.

\vspace{-1em}
\subsubsection{Integrable dynamics} 
Effectively, the parameter space $\mathscr C^*$ encodes the secular or `average' behaviour of the system. The dynamics can be easily determined from the transformed Hamiltonian $\mathcal H^*$ via the generalised Hamilton's equations \eqref{eq:hamilton_equations_bar}, without explicitly resorting to the variable transformations alluded to in the previous section.
We presently provide the expression of six independent osculating Keplerian parameters, following the bar-tilde conventions from Eq.\ \eqref{eq:convention_time_dependence}. The remaining parameters, such as $\bar v(t)$ and $\bar E(t)$, are obtained via the Keplerian relationships \eqref{eq:mean_anomaly}--\eqref{eq:true_and_eccentric_anomalies}, applied at each moment in time. The full set of parameters can then be used to reconstruct the evolution of the phase-space coordinates $(\vec r^*, \vec p^*)$ under the integrable Hamiltonian $\mathcal H^*$.
As expected, the secular PN perturbations act only on the orbital phase and the periapsis angle:
\begin{subequations}
\label{eq:parameters_evolution_bar}
\begin{align}
    \bar a(t) &= \bar a_0 \text, & 
    \bar e(t) &= \bar e_0\text, &
    \bar M(t)&= \bar M_0 + \overline{ \frac{\partial \mathcal H^*}{\partial L} } \, t \text, \\
    \bar \iota(t) &=\bar \iota_0 \text, & 
    \bar \Omega(t)&= \bar \Omega_0\text, & 
    \bar \varpi(t)&= \bar \varpi_0 + \overline{ \frac{\partial \mathcal H^*}{\partial J} } \, t \text.
\end{align}
\end{subequations}
The subscripted zeros refer to initial values obtained from the osculating functions applied to $(\vec r^*_0, \vec p^*_0)$.
Two of the constants, $\bar a_0$ and $\bar e_0$, are directly linked to the Noether quantities $\mathscr E$ and $\mathbfscr{J}$, which are also the dynamical invariants of the transformation $\mathcal T_g$. Namely, this symplectomorphism inherits both the time and the rotation symmetries of the system, as is clear from Eq.~\eqref{eq:generator_ADM}.
Rotational invariance in particular ensures the preservation of $\mathbfscr J$, so that $\vec J = \mathbfscr{J} = \vec {\bar J}$. Moreover, the energy relation $\mathscr E = \mathcal H^*(\bar{L}, \mathscr J)$ can be inverted order-by-order, yielding a useful expression:
\begin{align}
    \bar L =& \frac{1}{\sqrt{-2\mathscr E}} + \frac{1}{c^2}\Bigg\{ \frac{3}{\mathscr J} + \frac{\nu -15}{8}\sqrt{-2\mathscr E} \Bigg\} 
    + \frac{1}{c^4} \Bigg\{ \frac{35 - 10\nu}{4 \mathscr J^3} \nonumber\\& -\frac{6 \nu - 15}{2 \mathscr J}\mathscr E + \frac{3 \nu ^2+30 \nu +35}{128} (-2\mathscr E)^{3/2} \Bigg\} \text. \label{eq:bar_L}
\end{align}
The values for $\bar a_0$ and $\bar e_0$ can be deduced from these expressions by inputting $\bar a_0 = \bar L^2$ and $\bar e_0 = (1 - \bar J^2/ \bar L^2)^{1/2}$.

Next, we extract the precession and mean orbital frequencies from the normal-form Hamiltonian:
\begin{subequations}
    \begin{align}
        &\dot{\bar M}(t) = \frac{1}{\bar L^3} 
         + \frac{1}{c^2} \Bigg\{ \frac{9}{\bar J \bar L^4}+\frac{\nu -15}{2 \bar L^5} \Bigg\} 
         + \frac{1}{c^4} \Bigg\{ \frac{15 (7 - 2 \nu)}{4 \bar J^3 \bar L^4} \nonumber\\& \quad 
         +\frac{54}{\bar J^2 \bar L^5}+\frac{60 \nu - 525}{4 \bar J \bar L^6}+\frac{3 \left(\nu ^2-15 \nu +145\right)}{8 \bar L^7} \Bigg\} \text, \\
        &\dot{\bar \varpi}(t) = \frac{1}{c^2} \frac{3}{\bar J^2 \bar L^3} \nonumber \\& \quad\quad\;\;
        + \frac{1}{c^4} \Bigg\{ \frac{15 (7 - 2 \nu)}{4 \bar J^4 \bar L^3}+\frac{27}{\bar J^3 \bar L^4} +\frac{12 \nu - 105}{4 \bar J^2 \bar L^5}\Bigg\} \text.
    \end{align}
\end{subequations}
These frequencies are associated to well-known observables of the integrable PN dynamics. These are the periapsis-to-periapsis period $T$ and the periapsis advance per orbit $\Delta \phi$:
\begin{align}
    \dot{\bar M} &= G m \left(\frac{2\pi}{T}\right) \text, & \dot{\bar \varpi} &= \frac{\Delta \phi}{T} \text.
\end{align}
For convenience, we provide the observables $T$ and $\Delta \phi$ in physical units. Explicit expressions for $P$ and $\Delta \phi$ have also been derived in \cite{Damour1988, Damour2000a, Schafer2024} with the Hamilton-Jacobi framework. It is worth mentioning that their expressions perfectly align with our results when we incorporate Eq.~\eqref{eq:bar_L}.

\subsubsection{Complete dynamics of the original system} 
\label{sec:Keplerian_transformation}
We may now return to the original coordinate system $\mathscr C$. Applying $\mathcal T_g$ to each Keplerian parameter, we extract their full temporal evolution as a function of the secular variables derived in the previous section [cf.\ Eq.~\eqref{eq:Keplerian_parameter_transformation}]. 
After some lengthy bracket computations, we obtain the expressions that follow, strictly valid at $\mathcal O(c^{-6})$:
\begin{subequations}
\begin{align}
    \tilde a &= \bar a + \sum_{k=0}^6 \mathcal A^{(a)}_k \cos (k \bar v) \text, \label{eq:complete_dynamics_a} \\
    \tilde e &= \bar e + \sum_{k=0}^6 \mathcal A^{(e)}_k \cos (k \bar v) \text, 
\label{eq:complete_dynamics_e}\\
    \tilde v &= \bar v + \sum_{k=1}^6 \mathcal A^{(v)}_k \sin (k \bar v) + (\bar v - \bar E) \left( \sum_{k=0}^2 \mathcal B^{(v)}_k \cos (k \bar v) \right) \text, 
\label{eq:complete_dynamics_v}\\
    \tilde \varpi &= \bar \varpi + \sum_{k=1}^6 \mathcal A^{(\varpi)}_k \sin (k \bar v) + \mathcal B^{(\varpi)} (\bar v - \bar M) \text.
\label{eq:complete_dynamics_w}
\end{align}
\end{subequations}

The Kepler parameters $\tilde \iota = \bar \iota$ and $\tilde \Omega = \bar \Omega$ are identically unchanged by $\mathcal T_g$.
Each of the above coefficients is time-independent; they are reported in Appendix \ref{appendix:coefficients_starred_to_unstarred}. 

We remark that certain coefficients [Eqs.\ (\ref{eq:complete_dynamics_e}--\ref{eq:complete_dynamics_w})] contain coordinate singularities in the quasi-circular regime ($\bar e \to 0$ or equivalently $\bar J \to \bar L$). Specifically, their Laurent expansions contain single or double poles in $\bar e = 0$, and for small enough eccentricity ($\bar e \lesssim \varepsilon^2$, with $\varepsilon$ the small post-Newtonian parameter) the residuals are unbounded. This is an artefact of Taylor-expanding the osculating parametrisation in $\varepsilon$ and can be bypassed by adopting `Poincar\'e-like' elements:
\begin{align}
    \!\! z = e \exp{(\ii \varpi)} \text, && \lambda = v + \varpi \text, && \zeta = \sin(\iota/2) \exp(\ii \Omega) \text,
\end{align}

We have determined the full temporal evolution of the aforementioned `regularised' elements:
\begin{widetext}\vspace*{-5pt}
\begin{subequations}
\begin{align}
    \tilde z ={}& \bar z + \ee^{\ii \bar \varpi} \Bigg \{ 
    \sum_{k=-6}^6 \left(\mathcal A^{(z)}_k + \ii \, \mathcal B^{(z)}_k  \left(\bar v - \bar E \right) + \mathcal C^{(z)}_k (\bar v - \bar E)^2 \right)\ee^{\ii k (\bar \varpi - \bar \lambda)} 
    + \sum_{k=-2}^2 \left(\mathcal D^{(z)}_k + \ii \, \mathcal E^{(z)}_k  \left(\bar v - \bar E \right) \right)\ee^{\ii k (\bar \varpi - \bar {\lambda_E})}  \Bigg \}
    \text,\\
    \tilde \lambda ={}& \bar \lambda + \sum_{k=1}^4 \ii \, \mathcal A^{(\lambda)}_k \left( \bar z^k - (\bar z^\dagger)^k \right) \cos (k \bar \lambda) + \mathcal B^{(\lambda)}_k \left( \bar z^k + (\bar z^\dagger)^k \right) \sin (k \bar \lambda) \nonumber\\
    & + (\bar v - \bar E) \left( \sum_{k=0}^2 \mathcal C^{(\lambda)}_k \left( \bar z^k + (\bar z^\dagger)^k \right) \cos (k \bar \lambda) + \ii \, \mathcal D^{(\lambda)}_k \left( \bar z^k - (\bar z^\dagger)^k \right) \sin (k \bar \lambda) \right)
    + B^{(\varpi)} (\bar v - \bar M) \text, \\
    \tilde \zeta ={}& \bar \zeta \text,
\end{align}    
\end{subequations}
with $\lambda_E = E + \varpi$.
\vspace*{-1em} 
\end{widetext}

For convenience, the eccentricity can also be extracted via the exact expression instead of the truncated series:
\begin{equation}
    \tilde e = \sqrt{1- \bar J^2/{\tilde a}} \text, \label{eq:complete_dynamics_e_sqrt}
\end{equation}
with $\tilde a$ given by Eq.\ \eqref{eq:complete_dynamics_a}.\footnote{An expansion of Eq.\ \eqref{eq:complete_dynamics_e_sqrt} restores the truncated series \eqref{eq:complete_dynamics_e}. However, such expansion is not allowed when $\bar e \lesssim \varepsilon^2$ since the `small' parameter of the square-root exceeds its radius of convergence. Regardless, \eqref{eq:complete_dynamics_e_sqrt} remains valid for all values of $\bar e$.}

\subsubsection{Numerical example}

\begin{figure*}[!t]
    \centering
    \begin{tikzpicture}[x=0.95\linewidth,y=5cm]
        \draw (0, 0) node[inner sep=0] {\includegraphics[width=0.45\linewidth]{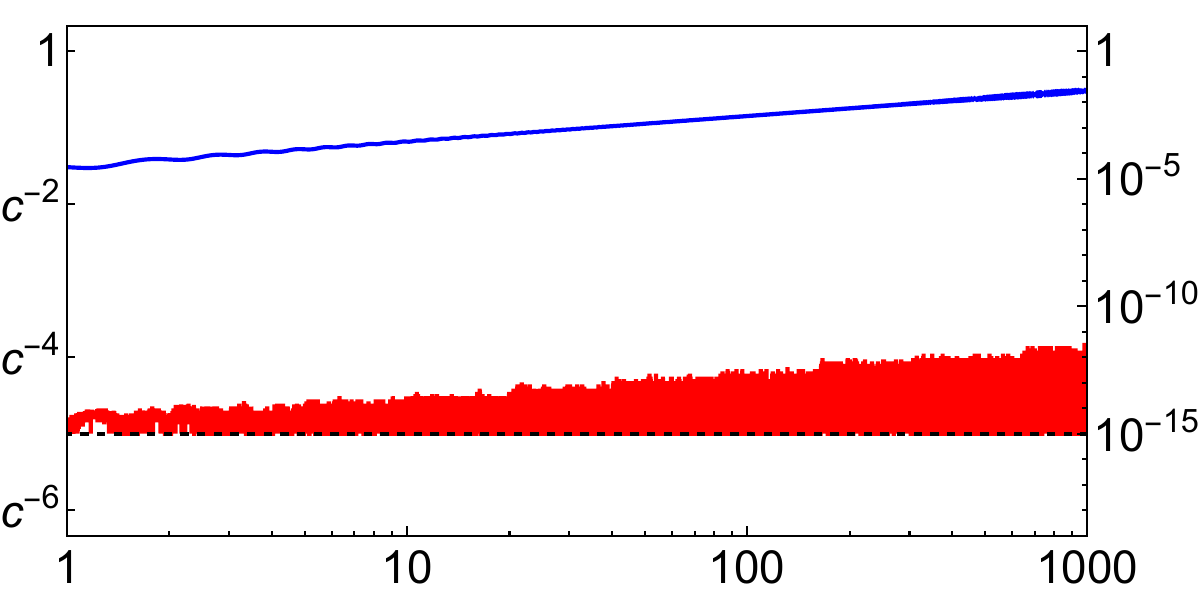}
        \includegraphics[width=0.45\linewidth]{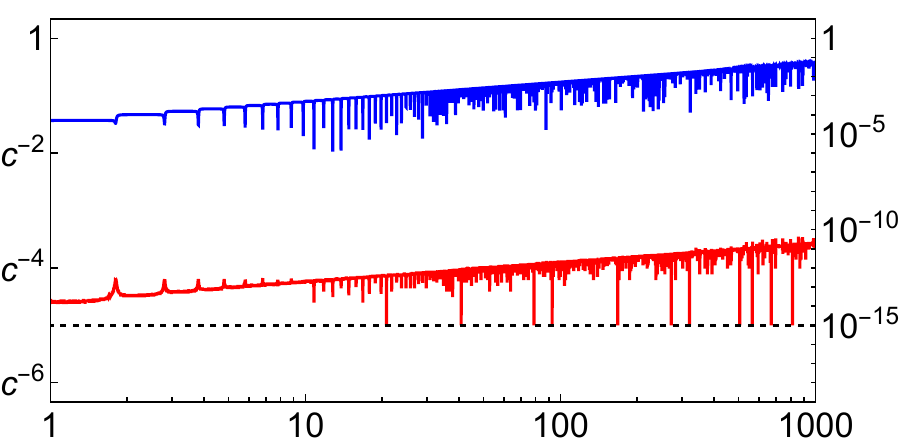}};
        \draw (0, -0.5) node{number of orbits};
        \draw (-0.5, -0) node[rotate=90] {phase residue $(\Delta \lambda)$};

        \draw (-0.443,0.3) node[draw=black, anchor=west] {$e = 0.01$};
        
        \draw (-0.452,0.33) node[fill=white, anchor=east] {$1\!$};
        \draw (-0.452,0.13) node[fill=white, anchor=east] {$\varepsilon^{\text{--} 2}\!\!$};
        \draw (-0.452,-0.08) node[fill=white, anchor=east] {$\varepsilon^{\text{--} 4}\!\!$};
        \draw (-0.452,-0.28) node[fill=white, anchor=east] {$\varepsilon^{\text{--} 6}\!\!$};
        \draw (-0.046, 0.33) node[fill=white, anchor=west] {$\!\!1$};
        \draw (-0.046, 0.165) node[fill=white, anchor=west] {$\!\!10^{\text{--} 5}$};
        \draw (-0.046, -0.005) node[fill=white, anchor=west] {$\!\!10^{\text{--} 10}$};
        \draw (-0.046, -0.18) node[fill=white, anchor=west] {$\!\!10^{\text{--} 15}$};

        \draw (-0.452, -0.325) node[fill=white, anchor=north] {$1$};
        \draw (-0.317, -0.325) node[fill=white, anchor=north] {$10$};
        \draw (-0.182, -0.325) node[fill=white, anchor=north] {$100$};
        \draw (-0.047, -0.325) node[fill=white, anchor=north] {$1000$};
        
        \draw (-0.06, 0.330) node[anchor=west] {-};
        \draw (-0.06, 0.159) node[anchor=west] {-};
        \draw (-0.06, -0.013) node[anchor=west] {-};
        \draw (-0.06, -0.1845) node[anchor=west] {-};
        
        \draw (+0.042,0.3) node[draw=black, anchor=west] {$e = 0.8$};
        
        \draw (0.028,0.33) node[fill=white, anchor=east] {$1\!$};
        \draw (0.028,0.13) node[fill=white, anchor=east] {$\varepsilon^{\text{--} 2}\!\!$};
        \draw (0.028,-0.08) node[fill=white, anchor=east] {$\varepsilon^{\text{--} 4}\!\!$};
        \draw (0.028,-0.28) node[fill=white, anchor=east] {$\varepsilon^{\text{--} 6}\!\!$};;
        \draw (0.435, 0.33) node[fill=white, anchor=west] {$\!\!1$};
        \draw (0.435, 0.165) node[fill=white, anchor=west] {$\!\!10^{\text{--} 5}$};
        \draw (0.435, -0.005) node[fill=white, anchor=west] {$\!\!10^{\text{--} 10}$};
        \draw (0.435, -0.175) node[fill=white, anchor=west] {$\!\!10^{\text{--} 15}$};
        
        \draw (0.03, -0.325) node[fill=white, anchor=north] {$1$};
        \draw (0.165, -0.325) node[fill=white, anchor=north] {$10$};
        \draw (0.3, -0.325) node[fill=white, anchor=north] {$100$};
        \draw (0.432, -0.325) node[fill=white, anchor=north] {$1000$};
        
        \draw (0.42, 0.331) node[anchor=west] {-};
        \draw (0.42, 0.160) node[anchor=west] {-};
        \draw (0.42, -0.012) node[anchor=west] {-};
        \draw (0.42, -0.1835) node[anchor=west] {-};
    \end{tikzpicture}
    \caption{Evolution of the phase residue ($\Delta \lambda = \lambda_{\mathrm{an}}^{2\mathrm{PN}} - \lambda_{\mathrm{num}}$; in \emph{solid red / light grey}) of the analytical solution truncated at 2PN with respect to numerical integration of the 2PN ADM Hamiltonian system [Eq.\ \eqref{eq:Hamiltonian_ADM}], for a 
    binary of perturbative parameter $\varepsilon \sim 10^{-3}$ and mass-ratio $\nu = 2/9$. Two eccentricity configurations are considered: on the \emph{left}, a quasi-circular orbit ($e = 0.01$); on the \emph{right}, a highly eccentric orbit ($e = 0.8$). The \emph{solid blue / dark grey} benchmark denotes the residual of a purely Keplerian analytical solution ($\lambda_{\mathrm{an}}^{0\mathrm{PN}} - \lambda_{\mathrm{num}}$). The numerical accuracy baseline at $10^{-15}$ is represented in \emph{dashed black}, estimated from the residue of the Keplerian solution when $\varepsilon \to \infty$. We have also verified that the conservation of the ADM constants $\mathscr E$ and $\mathscr J$ respect this accuracy level.}
    \label{fig:fast_motion_simulation}
\end{figure*}

\begin{figure*}[!ht]
    \centering
    \begin{minipage}{\linewidth}
     \begin{align*}
          && \mathcal{H}_\infty &\equiv \sum_{\ell=0}^{\infty} c^{-2\ell} \mathcal{H}_\ell &
          &\xrightarrow{\displaystyle \mathcal O(c^{-2K})} & 
          \mathcal{H} &\equiv \sum_{\ell=0}^{K-1} c^{-2\ell} \mathcal{H}_\ell &
          &\overset{\displaystyle \mathcal T_{g}}{\textcolor{red}{\xrightarrow{\hspace{3em}}}} & 
          \mathcal{H}^* &\equiv \mathcal T_g (\mathcal{H}) &&\\
          && &\quad\bigg\downarrow & && &\hspace{2em} \textcolor{blue}{\bigg\downarrow}& && &\textcolor{red}{\bigg\downarrow}& \\
          &&& \tilde x_\infty(t) &
          & \xleftarrow{{\displaystyle \mathcal O(c^{-2K})}} &
          \tilde x(t) &= \sum_{\ell = 0}^\infty c^{-2\ell} \tilde x_\ell(t) &
          & \overset{\displaystyle \mathcal T_{g}}{\textcolor{red}{\xleftarrow{\hspace{3em}}}} &
          &\bar x (t) &&
     \end{align*} 
     \end{minipage}%
     \caption{Schematics of the Lie canonical framework. The leftmost column conceptualises a complete relativistic Hamiltonian system with an infinite series of PN terms. The middle column represents a truncation at $\mathcal O(c^{-2K})$.
     The rightmost column corresponds to the secular system with Hamiltonian $\mathcal H^*$ in normal form, obtained by applying the Lie transformation $\mathcal T_g$ to $\mathcal{H}$.
     Indeed, the \emph{red / light grey} pathway illustrates the key transformation steps of the Lie approach; meanwhile, the \emph{blue / dark grey} arrow (\emph{middle column}) represents direct integration of $\mathcal H$, a nonlinear process with a potentially infinite amount of terms. In practical computations, each application of the Lie transform $\mathcal T_g$ is truncated to order $\mathcal O(c^{-2K})$, leading to a computation of $\tilde x(t)$ that is formally accurate to the same order.}
     \label{fig:coordinate_diagram}
\end{figure*}

The analytical expressions provided in the previous section solve the ADM Hamiltonian \eqref{eq:Hamiltonian_ADM} at the $\mathcal O(c^{-6})$ level. We shall now illustrate their validity numerically, for an unequal-mass moderately-relativistic binary in both quasi-circular and eccentric orbit configurations.
For each scenario, we integrate $\mathcal H$ by employing an 8th-order implicit Runge-Kutta scheme. Integration is performed in polar coordinates $(r, \lambda, p_r, J)$, where Hamilton's equations reduce to (see for instance \cite{LeTiec2015}; rotational invariance is required):

\null
\vspace*{-2em}
\begin{subequations}
    \begin{align}
    &&\dot r &= \frac{\partial \mathcal H}{\partial p_r} \text, &
    \dot p_r &= - \frac{\partial \mathcal H}{\partial r} \text, \\
    &&\dot \lambda &= \frac{\partial \mathcal H}{\partial J} \text, &
    \dot J &= - \frac{\partial \mathcal H}{\partial \lambda} = 0 \text,
    \end{align}
\end{subequations}
where an implicit change of variables is performed with the aid of Eqs.\ \eqref{eq:polynomial_r_subs}.
Figure~\ref{fig:fast_motion_simulation} depicts the phase residue ($\Delta \lambda = \lambda_{\mathrm{an}}^{\mathrm{2PN}} - \lambda_{\mathrm{num}}$; in $\mathscr C$ coordinate space) by which the 2PN perturbative solution $\lambda_{\mathrm{an}}^{\mathrm{2PN}}$ deviates from the numerical integration $\lambda_{\mathrm{num}}$ in each scenario.
As we might anticipate, this residue manifests close to 2PN marks below that of a standard Keplerian solution---that is, $\lambda_{\mathrm{an}}^{\mathrm{0PN}} - \lambda_{\mathrm{num}}$.
These residues are inevitable due to the truncation of higher-order terms in the Lie expansion of $\mathcal T_g$, as portrayed in Fig.~\ref{fig:coordinate_diagram}. 
Specifically, the secular Kepler parameters $\bar{x}(t)$ are obtained from a truncation of $\mathcal H^*$ to its integrable terms, while a second truncation occurs in the transformation $\tilde x(t) = \mathcal T_g (\bar x) (t)$.
Consequently, a residue manifests as terms of order $\mathcal O(c^{-6})$ that are linear in time for $\tilde \varpi(t)$ and $\tilde M(t)$ and periodic for the other variables, which is consistent with a 2PN solution.

\newpage
\section{Conclusions}
\label{sec:ccl}

We have provided a canonical Lie framework for analyzing the perturbed two-body problem under generic conditions of time-independence and rotational invariance. Our framework allows for the computation of the complete coordinate-time evolution in the PN conservative sector, using centre-of-mass coordinates $\mathscr C$. It is also applicable to more general perturbative two-body systems under similar conditions. The formalism can theoretically be applied to perturbations of arbitrary order, although the residue is expected to grow quickly at very high orders (see e.g. \cite{Morbidelli2002}). We provide a generic family of solutions [Eq.\ \eqref{eq:homologic_solution_iterative}] for systems within the aforementioned conditions. Accordingly, we specify an equivalent Hamiltonian in a new set of phase-space coordinates $\mathscr C^*$, which is integrable at given order; as well as the corresponding Lie transformation $\mathcal T_g : \mathscr C^* \to \mathscr C$, enabling the conversion between both coordinate systems. 
From the boundedness of the obtained Lie generator, it is clear that resonances are absent from the class of systems which have been studied. We have also applied our framework to the local, conservative ADM sector of the non-spinning point-mass binary problem, recovering classical results.

Going forward, the formalism may be broadened to other contexts, such as spinning pole-dipole models \cite{Damour2008, Barausse2009, Blanchet2013, Schafer2024} or time-dependent systems, including the ADM dissipative sector and non-local terms at 4PN \cite{Schafer2024}. Notably, the radiation-reaction term at $2.5$PN may be treated by either doubling the phase-space variables \cite{Buonanno1999, Galley2013, Galley2014, Schafer2024} or incorporating explicit time-dependence. In general, the presence of non-autonomous perturbative terms no longer leads to the preservation of the value of the conservative Hamiltonian by the Lie transform $\mathcal T_g$; regardlessly, similar homologic relations to Eq.~\eqref{eq:homologic_equation_system} can be derived, but in extended phase-space.
Spinning systems may benefit from the phase-space description of the Keplerian variables specified in Sec.~\ref{sec:Keplerian_parametrisation}; when properly normalised, the angular momentum and eccentricity vectors form the algebra of $SO(3) \times SO(3) \sim SO(4)$, potentially expressible in spinor form (see \cite{Deme2023} for an example). We conclude by remarking that in instances where resonances are present, order-by-order execution of the normal-form process is recommended (see e.g.\ \cite{Morbidelli2002} for details).



\begin{acknowledgments}
We are very grateful to G.\ Faye for the constructive comments and discussions.
C.A. acknowledges the joint finantial support of Centre National d'Études Spatiales (CNES) and École Doctorale Astronomie et Astrophysique d'Ile de France (ED127 AAIF). This work was also supported by the Programme National GRAM, by PNPS (CNRS/INSU), by INP and IN2P3 co-funded by CNES, and by CNES LISA grants at CEA/IRFU.
\end{acknowledgments}

\appendix

\section{Flow integrals}
\label{appendix:flow_integrals}
In this section we search for explicit expressions to the basis elements $\xi_k$ and $h_k$, which parametrise the solutions to the homologic equations.
We recall their definitions:
\begin{align*}
    \xi_k = \int \frac{1}{r^k} \left( \frac{\dd \theta}{\dd t}\right)^{-1}_{\mathrm{kep}} \,\dd \theta \text, && h_k = \avg{ \frac{1}{r^k} } \text,
\end{align*}
with the integration constant chosen such that $\avg{\xi_k} = 0$. It is immediately clear from the definition that $h_0 = 1$. Constraining $\theta$ to a Kepler angle further leads to $h_k = (n_0/2\pi) \big( \xi_k (\pi) - \xi_k (-\pi) \big)$, where $n_0 = 1/L^3$. For higher orders of $\xi_k$ and $h_k$ we shall split the computation into two distinct cases. Namely, using the definitions in Sec.\ \ref{sec:Keplerian_parametrisation}, we work out the relationships that follow:
\begin{itemize}
    \item For $k = 1$, identifying $\theta = E$, we have:
    \begin{align}
        \left( \frac{\dd E}{\dd t} \right)_{\mathrm{kep}} &= \frac{1}{L r} \text, & E = M + \frac{r p_r}{L} \text.\label{eq:eccentric_anomaly_phase_space}
    \end{align}    
    \item For $k \geq 2$, identifying $\theta = v$, we have instead:
    \begin{align}
        \left( \frac{\dd v}{\dd t} \right)_{\mathrm{kep}} &= \frac{J}{r^2} \text, &
        v &= M + \frac{r p_r}{L} + 2 \arctan\left( \frac{L r p_r}{J L + r} \right) \text. \label{eq:true_anomaly_phase_space}
        \end{align}
\end{itemize}
The integrals for $k = 1, 2$ are trivial and yield:
\begin{align}
    \xi_1 &= L E \text,  
    & h_1 &= L n_0 \text. \\
    \xi_2 &= \frac{v}{J} \text,  
    & h_2 &= \frac{n_0}{J} \text.
\end{align}
The evaluation of higher values of $k$ involves a little more work. Namely, we consider, for integer $k \geq 3$:
\begin{equation}
    \xi_{k} = \frac{1}{J^{2k-3}} \sum_{j=0}^{k-2} \binom{k-2}{j} e^j\int \cos^j{v} \,\dd v \text.
    \label{eq:flow_integral_2+i}
\end{equation}
Fortunately, the integrals of powers of cosine on the right $K_\alpha = \int \cos^\alpha v \, \dd v$ can be determined analytically. They follow from the recurrence relationship:
\begin{align}
    \!\!\!\! K_\alpha = \left(\frac{\alpha-1}{\alpha}\right)K_{\alpha-2} + \frac{1}{\alpha}\sin v \cos^{\alpha-1} v \text, && \! K_0 = v \text.
\end{align}
We verify via induction that the recurrence is satisfied~by:
\begin{equation}
    K_\alpha = \sin v \sum_{\beta \in \mathscr A} \frac{\beta!!}{\alpha!!} \frac{(\alpha - 1)!!}{(\beta - 1)!!} \frac{1}{\beta} \cos^{\beta-1}{v} + \alpha_{\mathrm{e}} \frac{(\alpha-1)!!}{\alpha!!} v \text,
\end{equation}
where $(\cdot)!!$ is the double factorial, $\alpha_{\mathrm{e}} \coloneqq 1 - (\alpha \bmod{2})$, and the sum is performed over the set $\mathscr A = \{ \alpha_{\mathrm{e}} + 1, \alpha_{\mathrm{e}} + 3, \ldots, \alpha - 2, \alpha\}$. The last term is only present when $\alpha$ is even, as evidenced by the presence of the $\alpha_{\mathrm{e}}$ factor.
Putting everything together and rearranging:
\begin{multline}
    \xi_k = \frac{p_r}{J^{2 k - 4}} \sum_{\alpha=0}^{k-2} \binom{k-2}{\alpha} \\ \times \sum_{\beta \in \mathscr A} \frac{\beta!!}{\alpha!!} \frac{(\alpha - 1)!!}{(\beta - 1)!!} \frac{e^{\alpha - \beta}}{\beta} \left( \frac{J^2}{r} - 1 \right)^{\beta - 1} \notag\\
    + \frac{v}{J^{2 k - 3}}\sum_{\alpha=0}^{\floor{\frac{k}{2}}-1} \binom{k-2}{2 \alpha} \binom{2\alpha}{\alpha} \left( \frac{e}{2} \right)^{2 \alpha} \text.
\end{multline}
\\ 

We can now extract $h_k$:
\begin{equation}
    h_{k} = \frac{n_0}{J^{2k - 3}}\sum_{\alpha=0}^{\floor{\frac{k}{2}}-1} \binom{k-2}{2\alpha}\binom{2\alpha}{\alpha} \left(\frac{e}{2}\right)^{2\alpha} \text,
\end{equation}
valid for integer $k \geq 2$, and where $\floor{y}$ is the largest integer smaller or equal to $y \in \mathbb R$. The angular relationships provided [Eqs.\ (\ref{eq:eccentric_anomaly_phase_space}, \ref{eq:true_anomaly_phase_space})] allow us to express $g_\ell = \sum_k A_{\ell, k} (\xi_k - (M/n_0) h_k)$ solely as a function of $r$, $p_r$ and $p$.
Finally, we can confirm via parity arguments that the average of each term does not contribute to $g_\ell$:
\begin{equation}
    \avg{\xi_k - \frac{M}{n_0} h_k} = 0 \text.
\end{equation}

\begin{widetext}
\section{Coefficients of the Keplerian coordinate transformation}
\label{appendix:coefficients_starred_to_unstarred}
We provide the coefficients of the 2PN ADM transformations for the Keplerian parameters from $\mathscr C^*$ to $\mathscr C$ as detailed in Sec.\ \ref{sec:Keplerian_transformation}:
\begin{itemize}
\item[$\bullet$] coefficients for a:
\begin{flalign}
    \mathcal A^{(a)}_0 ={}& \frac{1}{c^2} 
    \Bigg\{ 
        4-\nu -\frac{6 \bar L}{\bar J} +\frac{\bar L^2 (2 \nu -7)}{\bar J^2} + \frac{\bar L^4 (9-\nu )}{\bar J^4}
        + \frac{1}{c^2} \bigg(
            \frac{7 \left(7 \nu ^2-96 \nu +360\right) \bar{L}^6}{16 \bar{J}^8}
            -\frac{\left(207 \nu ^2-1876 \nu +4360\right) \bar{L}^4}{32 \bar{J}^6} \notag\\&
            +\frac{\left(17 \nu ^2-98 \nu +208\right) \bar{L}^2}{4 \bar{J}^4}
            +\frac{3 (7 \nu -90) \bar{L}^3}{4 \bar{J}^5} 
            -\frac{(\nu -32) \bar{L}}{4 \bar{J}^3}
            -\frac{3 (2 \nu -5)}{2 \bar{J} \bar{L}}
            -\frac{27 \nu ^2-196 \nu +40}{32 \bar{J}^2}
            +\frac{-\nu -16}{4 \bar{L}^2}
        \bigg)
    \Bigg\}  && \end{flalign} \begin{flalign}
    \mathcal A^{(a)}_1 ={}& \frac{\bar e}{c^2} \Bigg\{
        \frac{\bar L^2 (7 \nu -16)}{4 \bar J^2}
        +\frac{\bar L^4 (48-7 \nu )}{4 \bar J^4}
        + \frac{1}{c^2} \bigg(
        \frac{-24}{\bar{J} \left(\bar{J}+\bar{L}\right)}
        +\frac{640 + 124 \nu - 13 \nu ^2}{32 \bar{J}^2}
        -\frac{3 (9 \nu +40) \bar{L}}{8 \bar{J}^3} \\&   
        +\frac{\left(75 \nu ^2-426 \nu +1306\right) \bar{L}^2}{16 \bar{J}^4}
        +\frac{3 (35 \nu -312) \bar{L}^3}{8 \bar{J}^5}
        -\frac{\left(313 \nu ^2-2296 \nu +3684\right) \bar{L}^4}{32 \bar{J}^6}
        +\frac{\left(11 \nu ^2-147 \nu +504\right) \bar{L}^6}{2 \bar{J}^8}
        \bigg)
    \Bigg\} \text,  \notag\\
    \mathcal A^{(a)}_2 ={}& \frac{\bar e^2}{c^2} \Bigg\{
        \frac{\bar L^4 (3-\nu )}{\bar J^4}
        +\frac{1}{c^2} \bigg(   
            \frac{6 (\nu -7)}{\bar{J} \left(\bar{J}+\bar{L}\right)}
            -\frac{6 (\nu -7)}{\bar{J}^2}
            +\frac{6 (\nu -7) \bar{L}}{\bar{J}^3}
            +\frac{\left(67 \nu ^2-632 \nu +2120\right) \bar{L}^2}{32 \bar{J}^4}
            +\frac{3 (5 \nu -24) \bar{L}^3}{\bar{J}^5} \notag\\&
            -\frac{\left(105 \nu ^2-532 \nu +468\right) \bar{L}^4}{16 \bar{J}^6}
            +\frac{\left(127 \nu ^2-1536 \nu +4032\right) \bar{L}^6}{32 \bar{J}^8}
        \bigg)
    \Bigg\}  && \end{flalign} \begin{flalign}
    \mathcal A^{(a)}_3 ={}& \frac{\bar e}{c^2} \Bigg\{
        \frac{\bar L^2 \nu}{4\bar J^2} - \frac{\bar L^4 \nu}{4\bar J^4}
        +\frac{1}{c^2} \bigg( 
            \frac{-29 \nu ^2+200 \nu -256}{64 \bar{J}^2}
            -\frac{3 (11 \nu -16) \bar{L}}{16 \bar{J}^3}
            +\frac{\left(121 \nu ^2-710 \nu +1116\right) \bar{L}^2}{32 \bar{J}^4} \notag\\&
            +\frac{3 (55 \nu -144) \bar{L}^3}{16 \bar{J}^5}
            -\frac{\left(357 \nu ^2-2084 \nu +2744\right) \bar{L}^4}{64 \bar{J}^6}
            +\frac{3 \left(3 \nu ^2-29 \nu +48\right) \bar{L}^6}{4 \bar{J}^8}
        \bigg)
    \Bigg\}  && \end{flalign} \begin{flalign}
    \mathcal A^{(a)}_4 ={}& \frac{\bar e^2}{c^4} \Bigg\{
        \frac{(3 \nu -4) (11 \nu -48) \bar{L}^2}{32 \bar{J}^4}
        +\frac{3 (5 \nu -6) \bar{L}^3}{4 \bar{J}^5}
        -\frac{\left(63 \nu ^2-260 \nu +192\right) \bar{L}^4}{32 \bar{J}^6}
        +\frac{3 \left(5 \nu ^2-32 \nu +24\right) \bar{L}^6}{16 \bar{J}^8}
    \Bigg\}  && \end{flalign} \begin{flalign}
    \mathcal A^{(a)}_5 ={}& \frac{\bar e^2}{c^4} \Bigg\{
        \frac{3 \nu  (16- 3 \nu)}{64 \bar{J}^2}
        -\frac{9 \nu  \bar{L}}{16 \bar{J}^3}
        +\frac{\nu  (17 \nu -54) \bar{L}^2}{32 \bar{J}^4}
        +\frac{9 \nu  \bar{L}^3}{16 \bar{J}^5}
        -\frac{\nu  (41 \nu -108) \bar{L}^4}{64 \bar{J}^6}
        +\frac{(\nu -3) \nu  \bar{L}^6}{4 \bar{J}^8}
    \Bigg\}  && \end{flalign} \begin{flalign}
    \mathcal A^{(a)}_6 ={}& \frac{\bar e^2}{c^4} \Bigg\{
        \frac{\nu ^2 \bar{L}^2}{32 \bar{J}^4}
        -\frac{\nu ^2 \bar{L}^4}{16 \bar{J}^6}
        +\frac{\nu ^2 \bar{L}^6}{32 \bar{J}^8}
    \Bigg\} \text,
&& \end{flalign}

\item[$\bullet$] Coefficients for $e$:
\begin{flalign}
    \mathcal A^{(e)}_0 ={}& \frac{\bar{e}}{c^2} \Bigg\{ 
        \frac{3}{\bar{J} \bar{L}}
        -\frac{3}{\bar{J} \left(\bar{J}+\bar{L}\right)}
        -\frac{\nu -9}{2 \bar{J}^2}
        +\frac{\nu -4}{2 \bar{L}^2}
        + \frac{1}{c^2} \bigg(
            \frac{-181 \nu ^2+1144 \nu -496}{128 \bar{J}^2 \bar{L}^2}+\frac{2 \nu +13}{2 \bar{J}^3 \left(\bar{J}+\bar{L}\right)}+\frac{94-35 \nu }{8 \bar{J}^3 \bar{L}} \notag\\&
            +\frac{3 (8 \nu -29)}{4 \bar{J} \bar{L}^3}+\frac{2}{\bar{J}^3 \left(\bar{L}-\bar{J}\right)}-\frac{9}{2 \bar{J}^2 \left(\bar{J}+\bar{L}\right)^2}+\frac{157 \nu ^2-2296 \nu +5080}{128 \bar{J}^4}+\frac{3 \nu ^2-23 \nu +64}{8 \bar{L}^4}
        \bigg)
    \Bigg\} \text, 
&& \end{flalign}
\begin{flalign}
    \mathcal A^{(e)}_1 ={}& \frac{1}{c^2} \Bigg\{
        \frac{7 \nu -16}{8 \bar{L}^2}-\frac{7 \nu -48}{8 \bar{J}^2}
        + \frac{1}{c^2} \bigg(
            \frac{-89 \nu ^2+724 \nu -1174}{32 \bar{J}^2 \bar{L}^2}-\frac{9 (7 \nu -24)}{16 \bar{J}^3 \bar{L}}+\frac{9 (11 \nu -24)}{16 \bar{J} \bar{L}^3}+\frac{107 \nu ^2-1596 \nu +3516}{64 \bar{J}^4} \notag\\&
            +\frac{71 \nu ^2-404 \nu +640}{64 \bar{L}^4}
        \bigg)
    \Bigg\} \text,  
&& \end{flalign}
\begin{flalign}
    \mathcal A^{(e)}_2 ={}& \frac{\bar e}{c^2} \Bigg\{
        -\frac{(\nu -3)}{2 \bar{J}^2}
        +\frac{1}{c^2} \bigg( 
            \frac{-113 \nu ^2+352 \nu +32}{256 \bar{J}^2 \bar{L}^2}+\frac{3 \nu -29}{2 \bar{J}^3 \left(\bar{J}+\bar{L}\right)}-\frac{3}{2 \bar{J}^3 \bar{L}}-\frac{2}{\bar{J}^3 \left(\bar{L}-\bar{J}\right)} +\frac{49 \nu ^2-1632 \nu +3744}{256 \bar{J}^4}
        \bigg)
    \Bigg\} \text, 
&& \end{flalign}
\begin{flalign}
    \mathcal A^{(e)}_3 ={}& \frac{\bar e^2}{c^2} \Bigg\{
        \frac{\nu }{8 \bar{L}^2}-\frac{\nu }{8 \bar{J}^2}
        +\frac{1}{c^2} \bigg( 
            \frac{39 \nu ^2-340 \nu +828}{64 \bar{J}^2 \bar{L}^2}+\frac{3 (\nu +16)}{32 \bar{J} \bar{L}^3}+\frac{9 (13 \nu -48)}{32 \bar{J}^3 \bar{L}}-\frac{73 \nu ^2-224 \nu +248}{128 \bar{J}^4} \notag\\&
            +\frac{-5 \nu ^2+104 \nu -256}{128 \bar{L}^4}
        \bigg)
    \Bigg\} \text,  
&& \end{flalign}
\begin{flalign}
    \mathcal A^{(e)}_4 ={}& \frac{\bar e}{c^4} \Bigg\{
        \frac{45 \nu ^2-328 \nu +384}{128 \bar{J}^2 \bar{L}^2}+\frac{3 (5 \nu -6)}{8 \bar{J}^3 \bar{L}}-\frac{53 \nu ^2-168 \nu +168}{128 \bar{J}^4}
    \Bigg\} \text,  
&& \end{flalign}
\begin{flalign}
    \mathcal A^{(e)}_5 ={}& \frac{1}{c^4} \Bigg\{
        \frac{9 \nu }{32 \bar{J}^3 \bar{L}}+\frac{(11 \nu -36) \nu }{64 \bar{J}^2 \bar{L}^2}-\frac{9 \nu }{32 \bar{J} \bar{L}^3}-\frac{(13 \nu -24) \nu }{128 \bar{J}^4} -\frac{3 (3 \nu -16) \nu }{128 \bar{L}^4}
    \Bigg\} \text,&& \end{flalign}  \begin{flalign}
    \mathcal A^{(e)}_6 ={}& \frac{\bar e^3}{c^4} \Bigg\{  
        -\frac{\nu ^2}{256 \bar{J}^4}
    \Bigg\} \text,  
&& \end{flalign}

\item[$\bullet$] Coefficients for $v$:
\begin{flalign}    
    \mathcal A^{(v)}_1 ={}& \frac{1}{c^2} \Bigg\{
        -\frac{2}{\bar{J} \left(\bar{L}-\bar{J}\right)}-\frac{4}{\bar{J} \left(\bar{J}+\bar{L}\right)}+\frac{13 \nu -32}{8 \bar{J}^2}
        + \frac{1}{c^2} \bigg( 
            \frac{43 \nu ^2-260 \nu +1152}{64 \bar{J}^2 \bar{L}^2}+\frac{9 (23 \nu -72)}{16 \bar{J}^3 \bar{L}}+\frac{43 \nu -74}{16 \bar{J}^3 \left(\bar{L}-\bar{J}\right)}\notag\\&
            +\frac{178-41 \nu }{16 \bar{J}^3 \left(\bar{J}+\bar{L}\right)}-\frac{3}{\bar{J}^2 \left(\bar{J}+\bar{L}\right)^2}+\frac{\nu ^2+288 \nu -68}{64 \bar{J}^4}
        \bigg) 
    \Bigg\} \text,&& \end{flalign}  \begin{flalign}
    \mathcal A^{(v)}_2 ={}& \frac{1}{c^2} \Bigg\{
        -\frac{3}{2 \bar{J} \bar{L}}+\frac{\nu -2}{\bar{J}^2}+\frac{4-\nu }{2 \bar{L}^2}
        + \frac{1}{c^2} \bigg(
            \frac{-53 \nu ^2+688 \nu -1568}{128 \bar{J}^2 \bar{L}^2}-\frac{3 (14 \nu -47)}{8 \bar{J} \bar{L}^3}+\frac{79-8 \nu }{8 \bar{J}^3 \bar{L}}+\frac{4}{\bar{J}^3 \left(\bar{L}-\bar{J}\right)}\notag\\&
            -\frac{10}{\bar{J}^3 \left(\bar{J}+\bar{L}\right)}+\frac{117 \nu ^2-80 \nu +608}{128 \bar{J}^4}+\frac{-\nu ^2+7 \nu -32}{4 \bar{L}^4}
        \bigg)
    \Bigg\} \text,&& \end{flalign}  \begin{flalign} 
    \mathcal A^{(v)}_3 ={}& \frac{\bar e}{c^2} \Bigg\{     
        \frac{\nu}{8 \bar J^2}
        + \frac{1}{c^2} \bigg(          
            \frac{-67 \nu ^2+600 \nu -1280}{128 \bar{J}^2 \bar{L}^2}+\frac{3-\nu }{\bar{J}^3 \left(\bar{L}-\bar{J}\right)}-\frac{3 (41 \nu -176)}{32 \bar{J}^3 \bar{L}}+\frac{-19 \nu -36}{8 \bar{J}^3 \left(\bar{J}+\bar{L}\right)}\notag\\&
            +\frac{163 \nu ^2-580 \nu +792}{128 \bar{J}^4}
        \bigg)
    \Bigg\} \text,&& \end{flalign}  \begin{flalign} 
    \mathcal A^{(v)}_4 ={}& \frac{1}{c^4} \Bigg\{     
        \frac{-21 \nu ^2+126 \nu -92}{32 \bar{J}^2 \bar{L}^2}+\frac{3 (\nu -4)}{4 \bar{J} \bar{L}^3}-\frac{3 (7 \nu -8)}{8 \bar{J}^3 \bar{L}}+\frac{(\nu -2) (21 \nu -32)}{32 \bar{J}^4}+\frac{(\nu -4)^2}{8 \bar{L}^4}
    \Bigg\} \text,&& \end{flalign}  \begin{flalign} 
    \mathcal A^{(v)}_5 ={}& \frac{\bar e}{c^4} \Bigg\{     
        -\frac{9 \nu }{32 \bar{J}^3 \bar{L}}-\frac{3 (3 \nu -16) \nu }{128 \bar{J}^2 \bar{L}^2}+\frac{(17 \nu -36) \nu }{128 \bar{J}^4}
    \Bigg\} \text,&& \end{flalign}  \begin{flalign} 
    \mathcal A^{(v)}_6 ={}& \frac{\bar e^2}{c^4} \Bigg\{   
        \frac{\nu ^2}{128 \bar J^4}
    \Bigg\} \text,&& \end{flalign}  \begin{flalign}
    \mathcal B^{(v)}_0 ={}& \frac{1}{c^4}\Bigg\{
        \frac{9 (2 \nu -5)}{4 \bar{J}^4}-\frac{3 (2 \nu -5)}{4 \bar{J}^2 \bar{L}^2}
    \Bigg\} \text,&& \end{flalign}  \begin{flalign}
    \mathcal B^{(v)}_1 ={}& \frac{\bar e}{c^4}\Bigg\{ 
        \frac{3 (2 \nu -5)}{\bar{J}^4}
    \Bigg\} \text,&& \end{flalign}  \begin{flalign}
    \mathcal B^{(v)}_2 ={}& \frac{\bar e^2}{c^4}\Bigg\{ 
        \frac{3 (2 \nu -5)}{4 \bar{J}^4}
    \Bigg\} \text. 
&& \end{flalign}

\item[$\bullet$] Coefficients for $\varpi$:
\begin{flalign}
        \mathcal A^{(\varpi)}_1 ={}& \frac{\bar e}{c^2} \Bigg\{ 
            \frac{2}{\bar{J} \left(\bar{L}-\bar{J}\right)}-\frac{2}{\bar{J} \left(\bar{J}+\bar{L}\right)}-\frac{5 \nu -48}{8 \bar{J}^2}
            + \frac{1}{c^2} \bigg( 
                \frac{-7 \nu ^2+28 \nu -128}{64 \bar{J}^2 \bar{L}^2}-\frac{9 (3 \nu -8)}{16 \bar{J}^3 \bar{L}}+\frac{74-43 \nu }{16 \bar{J}^3 \left(\bar{L}-\bar{J}\right)} \notag\\&
                +\frac{25 \nu -362}{16 \bar{J}^3 \left(\bar{J}+\bar{L}\right)}-\frac{6}{\bar{J}^2 \left(\bar{J}+\bar{L}\right)^2}+\frac{11 \nu ^2-712 \nu +2100}{64 \bar{J}^4}
            \bigg)
        \Bigg\} \text,&& \end{flalign}  \begin{flalign}
        \mathcal A^{(\varpi)}_2 ={}& \frac{1}{c^2} \Bigg\{ 
            -\frac{\nu -3}{2 \bar{J}^2}
            + \frac{1}{c^2} \bigg( 
                \frac{21 \nu ^2-304 \nu +1296}{128 \bar{J}^2 \bar{L}^2}+\frac{3 (\nu -14)}{4 \bar{J}^3 \bar{L}}-\frac{4}{\bar{J}^3 \left(\bar{L}-\bar{J}\right)}-\frac{8}{\bar{J}^3 \left(\bar{J}+\bar{L}\right)}-\frac{53 \nu ^2-176 \nu +336}{128 \bar{J}^4}
            \bigg)
        \Bigg\} \text,&& \end{flalign}  \begin{flalign}
        \mathcal A^{(\varpi)}_3 ={}& \frac{\bar e}{c^2} \Bigg\{ 
            -\frac{\nu}{8 \bar J^2}
            + \frac{1}{c^2} \bigg( 
                \frac{11 \nu ^2-104 \nu +256}{128 \bar{J}^2 \bar{L}^2}+\frac{3 (\nu -16)}{32 \bar{J}^3 \bar{L}}+\frac{\nu -3}{\bar{J}^3 \left(\bar{L}-\bar{J}\right)}+\frac{19 \nu -72}{8 \bar{J}^3 \left(\bar{J}+\bar{L}\right)}-\frac{107 \nu ^2-564 \nu +824}{128 \bar{J}^4}
            \bigg)
        \Bigg\} \text,&& \end{flalign}  \begin{flalign}
        \mathcal A^{(\varpi)}_4 ={}& \frac{1}{c^4} \Bigg\{
            \frac{13 \nu ^2-86 \nu +96}{32 \bar{J}^2 \bar{L}^2}+\frac{3 (5 \nu -6)}{8 \bar{J}^3 \bar{L}}-\frac{17 \nu ^2-66 \nu +60}{32 \bar{J}^4}
        \Bigg\} \text,&& \end{flalign}  \begin{flalign} 
        \mathcal A^{(\varpi)}_5 ={}& \frac{\bar e}{c^4} \Bigg\{
            \frac{9 \nu }{32 \bar{J}^3 \bar{L}}+\frac{3 (3 \nu -16) \nu }{128 \bar{J}^2 \bar{L}^2}-\frac{(17 \nu -36) \nu }{128 \bar{J}^4}
        \Bigg\} \text,&& \end{flalign}  \begin{flalign} 
        \mathcal A^{(\varpi)}_6 ={}& \frac{\bar e^2}{c^4} \Bigg\{
            -\frac{\nu ^2}{128 \bar{J}^4}
        \Bigg\} \text,&& \end{flalign}  \begin{flalign} 
        \mathcal B^{(\varpi)} ={}&
        \frac{1}{c^2} \Bigg\{
            \frac{3}{\bar J^2}
            + \frac{1}{c^2} \bigg( 
                \frac{3 (2 \nu -5)}{4 \bar{J}^2 \bar{L}^2}-\frac{15 (2 \nu -7)}{4 \bar{J}^4}
            \bigg)
        \Bigg\} \text.
&& \end{flalign}
        
\item[$\bullet$] The non-zero coefficients for $z$ are:
\begin{flalign}
    \mathcal A^{(z)}_0 ={}& 
    \frac{\bar e}{c^2} \Bigg\{ 
        \frac{3}{\bar{J} \bar{L}}
        -\frac{3}{\bar{J} \left(\bar{J}
        +\bar{L}\right)}-\frac{\nu -9}{2 \bar{J}^2}
        +\frac{\nu -4}{2 \bar{L}^2}
        + \frac{1}{c^2} \bigg( 
            \frac{-21 \nu ^2+133 \nu -10}{16 \bar{J}^2 \bar{L}^2}
            -\frac{29 (\nu -2)}{8 \bar{J}^3 \bar{L}}
            +\frac{3 (8 \nu -29)}{4 \bar{J} \bar{L}^3}
            +\frac{2 \nu +17}{2 \bar{J}^3 \left(\bar{J}
            +\bar{L}\right)} \notag\\&
            -\frac{9}{2 \bar{J}^2 \left(\bar{J}
            +\bar{L}\right)^2} 
            +\frac{17 \nu ^2-251 \nu +446}{16 \bar{J}^4}
            +\frac{3 \nu ^2-23 \nu +64}{8 \bar{L}^4}
        \bigg) 
    \Bigg\}  && \end{flalign} \begin{flalign}
    \mathcal A^{(z)}_1 ={}& 
    \frac{1}{c^2} \Bigg\{ 
        -\frac{\nu \bar e^2}{8 \bar{J}^2}
        + \frac{1}{c^2} \bigg(
            \frac{-13 \nu ^2+63 \nu -149}{16 \bar{J}^2 \bar{L}^2}-\frac{3 (\nu -72)}{16 \bar{J}^3 \bar{L}}+\frac{3 (5 \nu -8)}{16 \bar{J} \bar{L}^3}+\frac{15 \nu ^2-92 \nu -150}{32 \bar{J}^4}+\frac{11 \nu ^2-58 \nu +64}{32 \bar{L}^4}
        \bigg) 
    \Bigg\}  && \end{flalign} \begin{flalign}
    \mathcal A^{(z)}_{-1} ={}& 
    \frac{1}{c^2} \Bigg\{ 
        \frac{3 (8 - \nu) \bar e^2}{4 \bar{J}^2}
        + \frac{1}{c^2} \bigg(
            -\frac{3 \left(19 \nu ^2-188 \nu +276\right)}{32 \bar{J}^2 \bar{L}^2}-\frac{9 (3 \nu -8)}{8 \bar{J}^3 \bar{L}}+\frac{3 (17 \nu -48)}{8 \bar{J} \bar{L}^3}+\frac{65 \nu ^2-1280 \nu +3528}{64 \bar{J}^4}\notag\\&
            +\frac{49 \nu ^2-288 \nu +512}{64 \bar{L}^4}
        \bigg) 
    \Bigg\} \text,&& \end{flalign}  \begin{flalign} 
    \mathcal A^{(z)}_2 ={}& 
    \frac{\bar e}{c^4} \Bigg\{ 
        \frac{-\nu ^2+8 \nu -48}{16 \bar{J}^2 \bar{L}^2}+\frac{3 (\nu +6)}{8 \bar{J}^3 \bar{L}}+\frac{\nu ^2-14 \nu +12}{16 \bar{J}^4}
    \Bigg\}  && \end{flalign} \begin{flalign}
    \mathcal A^{(z)}_{-2} ={}& 
    \frac{\bar e}{c^2} \Bigg\{ 
        -\frac{\nu -3}{2 \bar{J}^2}
        + \frac{1}{c^2} \bigg(
            \frac{-7 \nu ^2+22 \nu +34}{16 \bar{J}^2 \bar{L}^2}+\frac{3 (\nu -11)}{2 \bar{J}^3 \left(\bar{J}+\bar{L}\right)}-\frac{3 (3 \nu -2)}{8 \bar{J}^3 \bar{L}}+\frac{3 \nu ^2-112 \nu +366}{16 \bar{J}^4}
        \bigg) 
    \Bigg\}  && \end{flalign} \begin{flalign}
    \mathcal A^{(z)}_3 ={}& 
    \frac{1}{c^4} \Bigg\{ 
        \frac{3 \nu }{32 \bar{J}^3 \bar{L}}+\frac{(\nu -10) \nu }{64 \bar{J}^2 \bar{L}^2}-\frac{3 \nu }{32 \bar{J} \bar{L}^3}-\frac{(\nu -4) \nu }{128 \bar{J}^4}-\frac{(\nu -16) \nu }{128 \bar{L}^4}
    \Bigg\}  && \end{flalign} \begin{flalign}
    \mathcal A^{(z)}_{-3} ={}& 
    \frac{1}{c^2} \Bigg\{ 
        \frac{\nu }{8 \bar{L}^2}-\frac{\nu }{8 \bar{J}^2}
        + \frac{1}{c^2} \bigg(
            \frac{7 \nu ^2-67 \nu +183}{16 \bar{J}^2 \bar{L}^2}+\frac{3 (3 \nu +8)}{16 \bar{J} \bar{L}^3}+\frac{3 (17 \nu -72)}{16 \bar{J}^3 \bar{L}}-\frac{13 \nu ^2+8 \nu -82}{32 \bar{J}^4}+\frac{-\nu ^2+22 \nu -64}{32 \bar{L}^4}
        \bigg) 
    \Bigg\}  && \end{flalign} \begin{flalign}
    \mathcal A^{(z)}_{-4} ={}& 
    \frac{\bar e}{c^2} \Bigg\{ 
        \frac{5 \nu ^2-39 \nu +48}{16 \bar{J}^2 \bar{L}^2}+\frac{3 (5 \nu -6)}{8 \bar{J}^3 \bar{L}}-\frac{5 \nu ^2-9 \nu +12}{16 \bar{J}^4} 
    \Bigg\}  && \end{flalign} \begin{flalign}
    \mathcal A^{(z)}_{-5} ={}& 
    \frac{1}{c^4} \Bigg\{ 
        \frac{9 \nu }{32 \bar{J}^3 \bar{L}}+\frac{3 (3 \nu -10) \nu }{64 \bar{J}^2 \bar{L}^2}-\frac{9 \nu }{32 \bar{J} \bar{L}^3}-\frac{3 (3 \nu -4) \nu }{128 \bar{J}^4}-\frac{3 (3 \nu -16) \nu }{128 \bar{L}^4}
    \Bigg\}  && \end{flalign} \begin{flalign}
    \mathcal B^{(z)}_0 ={}& 
    \frac{ \bar e}{c^2} \Bigg\{ 
        \frac{3}{\bar{J}^2}
        + \frac{1}{c^2} \bigg( 
            \frac{3 (4 \nu -13)}{4 \bar{J}^2 \bar{L}^2}+\frac{9}{\bar{J}^3 \bar{L}}-\frac{9}{\bar{J}^3 \left(\bar{J}+\bar{L}\right)}-\frac{3 (12 \nu -53)}{4 \bar{J}^4}
        \bigg) 
    \Bigg\} \text,&& \end{flalign}  \begin{flalign} 
    \mathcal B^{(z)}_1 ={}& 
    \frac{ \bar e^2}{c^4} \Bigg\{ 
        -\frac{3 \nu }{8 \bar{J}^4}
    \Bigg\} \text,&& \end{flalign}  \begin{flalign}
    \mathcal B^{(z)}_{-1} ={}& 
    \frac{1}{c^4} \Bigg\{ 
        \frac{3 (3 \nu -8)}{4 \bar{J}^2 \bar{L}^2}-\frac{9 (\nu -8)}{4 \bar{J}^4}
    \Bigg\}  && \end{flalign} \begin{flalign}
    \mathcal B^{(z)}_{-2} ={}& 
    \frac{ \bar e}{c^4} \Bigg\{ 
        -\frac{3 (\nu -3)}{2 \bar{J}^4}
    \Bigg\}  && \end{flalign} \begin{flalign}
    \mathcal B^{(z)}_{-3} ={}& 
    \frac{ \bar e^2}{c^4} \Bigg\{ 
        -\frac{3 \nu }{8 \bar{J}^4}
    \Bigg\}  && \end{flalign} \begin{flalign}
    \mathcal C^{(z)}_{0} ={}& 
    \frac{1}{c^4} \Bigg\{ 
        -\frac{9}{2 \bar{J}^4}
    \Bigg\}  && \end{flalign} \begin{flalign}
    \mathcal D^{(z)}_1 ={}& 
    \frac{1}{c^2} \Bigg\{ 
        -\frac{3 \bar e^2}{2 \bar{J}^2}
        + \frac{1}{c^2} \bigg(
            -\frac{3 (\nu -10)}{4 \bar{J} \bar{L}^3}-\frac{3 (3 \nu -10)}{2 \bar{J}^2 \bar{L}^2}-\frac{9}{2 \bar{J}^3 \bar{L}}+\frac{15 (2 \nu -7)}{8 \bar{J}^4}+\frac{3 (4 \nu -13)}{8 \bar{L}^4}
        \bigg) 
    \Bigg\} \text,&& \end{flalign}  \begin{flalign} 
    \mathcal D^{(z)}_{-1} ={}& 
    \frac{1}{c^2} \Bigg\{ 
        \frac{3 \bar e^2}{2 \bar{J}^2}
        + \frac{1}{c^2} \bigg(
            -\frac{3 (\nu +2)}{4 \bar{J} \bar{L}^3}+\frac{3 (3 \nu -10)}{2 \bar{J}^2 \bar{L}^2}-\frac{9}{2 \bar{J}^3 \bar{L}}-\frac{15 (2 \nu -7)}{8 \bar{J}^4}-\frac{3 (4 \nu -13)}{8 \bar{L}^4}
        \bigg) 
    \Bigg\} \text,&& \end{flalign}  \begin{flalign} 
    \mathcal D^{(z)}_2 ={}& 
    \frac{\bar e^3}{c^4} \Bigg\{ 
        \frac{9}{8 \bar{J}^4}
    \Bigg\} \text,&& \end{flalign}  \begin{flalign} 
    \mathcal D^{(z)}_{-2} ={}& 
    \frac{\bar e^3}{c^4} \Bigg\{ 
        \frac{9}{8 \bar{J}^4}
    \Bigg\} \text,&& \end{flalign}  \begin{flalign} 
    \mathcal E^{(z)}_1 ={}& 
    \frac{1}{c^2} \Bigg\{ 
        -\frac{9}{2 \bar{J}^4}
    \Bigg\} \text,&& \end{flalign}  \begin{flalign}
    \mathcal E^{(z)}_{-1} ={}& 
    \frac{1}{c^2} \Bigg\{ 
        \frac{9}{2 \bar{J}^4}
    \Bigg\}  
&& \end{flalign}
The remaining coefficients are identically zero.

\item[$\bullet$] The non-zero coefficients for $\lambda$ are:
\begin{flalign}
        \mathcal A^{(\lambda)}_1 ={}& 
        \frac{1}{c^2} \Bigg\{ 
            \frac{(\nu +2)}{2 J^2}-\frac{3}{J (J+L)}
            + \frac{1}{c^2} \bigg( 
                \frac{\left(3 \nu ^2-106 \nu +508\right)}{32 J^4}
                -\frac{(2 \nu +23)}{4 J^3 (J+L)}
                +\frac{9 (5 \nu -16)}{8 J^3 L}
                +\frac{\left(9 \nu ^2-58 \nu +256\right)}{32 J^2 L^2}\notag\\&
                -\frac{9}{2 J^2 (J+L)^2}
            \bigg) 
        \Bigg\}  && \end{flalign} \begin{flalign}    
        \mathcal A^{(\lambda)}_2 ={}& 
        \frac{1}{c^2} \Bigg\{ 
            \frac{(\nu -1)}{4 J^2}-\frac{3}{4 J (J+L)}
            + \frac{1}{c^2} \bigg( 
                \frac{\left(4 \nu ^2+6 \nu +17\right)}{16 J^4}
                -\frac{(11 \nu +2)}{4 J^3 (J+L)}
                +\frac{3 (14 \nu -47)}{16 J^3 L}
                +\frac{\left(\nu ^2-7 \nu +32\right)}{8 J^2 L^2}\notag\\&
                +\frac{9}{2 J^2 (J+L)^2}
            \bigg) 
        \Bigg\}  && \end{flalign} \begin{flalign}    
        \mathcal A^{(\lambda)}_3 ={}& 
        \frac{1}{c^4} \Bigg\{ 
            \frac{\left(7 \nu ^2-2 \nu -4\right)}{32 J^4}
            -\frac{3 (5 \nu -2)}{8 J^3 (J+L)}
            +\frac{27}{8 J^2 (J+L)^2}
        \Bigg\}  && \end{flalign} \begin{flalign}    
        \mathcal A^{(\lambda)}_4 ={}& 
        \frac{1}{c^4} \Bigg\{ 
            \frac{(\nu -1)^2}{16 J^4}
            -\frac{3 (\nu -1)}{8 J^3 (J+L)}
            +\frac{9}{16 J^2 (J+L)^2}
        \Bigg\}  && \end{flalign} \begin{flalign}    
        \mathcal B^{(\lambda)}_1 ={}& 
        \frac{1}{c^2} \Bigg\{ 
            \frac{\nu +2}{2 J^2}-\frac{3}{J (J+L)}
            + \frac{1}{c^2} \bigg( 
                \frac{3 \nu ^2-106 \nu +508}{32 J^4}+\frac{9 (5 \nu -16)}{8 J^3 L}+\frac{-2 \nu -23}{4 J^3 (J+L)}+\frac{9 \nu ^2-58 \nu +256}{32 J^2 L^2}\notag\\&
                -\frac{9}{2 J^2 (J+L)^2}
            \bigg) 
        \Bigg\}  && \end{flalign} \begin{flalign}
        \mathcal B^{(\lambda)}_2 ={}& 
        \frac{1}{c^2} \Bigg\{ 
            \frac{\nu -1}{4 J^2}-\frac{3}{4 J (J+L)}
            + \frac{1}{c^2} \bigg( 
                \frac{4 \nu ^2+6 \nu +17}{16 J^4}+\frac{3 (14 \nu -47)}{16 J^3 L}+\frac{-11 \nu -2}{4 J^3 (J+L)}+\frac{\nu ^2-7 \nu +32}{8 J^2 L^2}+\frac{9}{2 J^2 (J+L)^2}
            \!\bigg)\! 
        \Bigg\}  && \end{flalign} \begin{flalign}
        \mathcal B^{(\lambda)}_3 ={}& 
        \frac{1}{c^4} \Bigg\{ 
            \frac{7 \nu ^2-2 \nu -4}{32 J^4}-\frac{3 (5 \nu -2)}{8 J^3 (J+L)}+\frac{27}{8 J^2 (J+L)^2}
        \Bigg\}  && \end{flalign} \begin{flalign}
        \mathcal B^{(\lambda)}_4 ={}& 
        \frac{1}{c^4} \Bigg\{ 
            \frac{(\nu -1)^2}{16 J^4}-\frac{3 (\nu -1)}{8 J^3 (J+L)}+\frac{9}{16 J^2 (J+L)^2}
        \Bigg\}  && \end{flalign} \begin{flalign}
        \mathcal C^{(\lambda)}_0 ={}& 
        \frac{1}{c^4} \Bigg\{ 
            \frac{15 (2 \nu -7)}{4 J^4} - \frac{3 (\nu -5)}{J^4} - \frac{3 (2 \nu -5)}{4 J^2 L^2}
        \Bigg\}  && \end{flalign}  \begin{flalign}
        \mathcal C^{(\lambda)}_1 ={}& \mathcal C^{(\lambda)}_2 = 
        \frac{1}{c^4} \Bigg\{ 
            \frac{3 (2 \nu -5)}{2 J^4}
        \Bigg\} 
&& \end{flalign} 
\begin{flalign}
        \mathcal D^{(\lambda)}_1 ={}& \mathcal D^{(\lambda)}_2 =
        \frac{1}{c^4} \Bigg\{ 
            -\frac{3 (2 \nu -5)}{2 J^4}
        \Bigg\}  
&& \end{flalign}

The remaining coefficients are identically zero.

\end{itemize}
\end{widetext}

\bibliography{references}

\end{document}